\documentclass[longauth]{aa}

\usepackage{graphicx}
\usepackage{natbib}
\usepackage{scalerel}

\usepackage[table]{xcolor}
\usepackage{soul}

\newcommand{\lcal}{ {\mathcal L} }
\newcommand{\pcal}{ {\mathcal P} }
\newcommand{\mcal}{ {\mathcal M} }
\newcommand{\ccal}{ {\mathcal C} }
\newcommand{\dif}{ {\mathrm d} }
\newcommand{\deus}{\texttt{DEUS-PUR}}

\newcommand\mr{\mathrm}

\newcommand\invMpc{h/\mr{Mpc}}
\newcommand\Mpc{\mr{Mpc}/h}

\newcommand\invMpcc{(h/\mr{Mpc})^3}
\newcommand\Gpcc{(\mr{Gpc}/h)^3}

\newcommand{\covmos}{\texttt{COVMOS}}
\newcommand{\demcov}{\texttt{DEMNUni-Cov}}

\defcitealias{KP2}{GB-2025}

\usepackage{txfonts}
\usepackage[pdfencoding=auto,psdextra]{hyperref}
\hypersetup{
    colorlinks=true,
    linkcolor=blue,
    filecolor=magenta,      
    urlcolor=blue,
    citecolor=blue
}
\makeatletter
\renewcommand*\aa@pageof{, page \thepage{} of 19}
\makeatother

\usepackage[utf8]{inputenc}

\usepackage[switch, modulo]{lineno}

\usepackage{euclid}

\begin{document}

\newcommand{\orcid}[1]{} 
\author{Euclid Collaboration: J.~Bel\thanks{\email{jbel@cpt.univ-mrs.fr}}\inst{\ref{aff1}}
\and S.~Gouyou~Beauchamps\inst{\ref{aff2},\ref{aff3}}
\and P.~Baratta\orcid{0000-0001-5533-8437}\inst{\ref{aff4},\ref{aff5}}
\and L.~Blot\orcid{0000-0002-9622-7167}\inst{\ref{aff6},\ref{aff7}}
\and C.~Carbone\orcid{0000-0003-0125-3563}\inst{\ref{aff8}}
\and P.-S.~Corasaniti\orcid{0000-0002-6386-7846}\inst{\ref{aff7},\ref{aff9}}
\and E.~Sefusatti\orcid{0000-0003-0473-1567}\inst{\ref{aff10},\ref{aff11},\ref{aff12}}
\and S.~Escoffier\orcid{0000-0002-2847-7498}\inst{\ref{aff4}}
\and W.~Gillard\orcid{0000-0003-4744-9748}\inst{\ref{aff4}}
\and A.~Amara\inst{\ref{aff13}}
\and S.~Andreon\orcid{0000-0002-2041-8784}\inst{\ref{aff14}}
\and N.~Auricchio\orcid{0000-0003-4444-8651}\inst{\ref{aff15}}
\and C.~Baccigalupi\orcid{0000-0002-8211-1630}\inst{\ref{aff11},\ref{aff10},\ref{aff12},\ref{aff16}}
\and M.~Baldi\orcid{0000-0003-4145-1943}\inst{\ref{aff17},\ref{aff15},\ref{aff18}}
\and S.~Bardelli\orcid{0000-0002-8900-0298}\inst{\ref{aff15}}
\and P.~Battaglia\orcid{0000-0002-7337-5909}\inst{\ref{aff15}}
\and A.~Biviano\orcid{0000-0002-0857-0732}\inst{\ref{aff10},\ref{aff11}}
\and E.~Branchini\orcid{0000-0002-0808-6908}\inst{\ref{aff19},\ref{aff20},\ref{aff14}}
\and M.~Brescia\orcid{0000-0001-9506-5680}\inst{\ref{aff21},\ref{aff22}}
\and J.~Brinchmann\orcid{0000-0003-4359-8797}\inst{\ref{aff23},\ref{aff24},\ref{aff25}}
\and S.~Camera\orcid{0000-0003-3399-3574}\inst{\ref{aff26},\ref{aff27},\ref{aff28}}
\and G.~Ca\~nas-Herrera\orcid{0000-0003-2796-2149}\inst{\ref{aff29},\ref{aff30},\ref{aff31}}
\and V.~Capobianco\orcid{0000-0002-3309-7692}\inst{\ref{aff28}}
\and V.~F.~Cardone\inst{\ref{aff32},\ref{aff33}}
\and J.~Carretero\orcid{0000-0002-3130-0204}\inst{\ref{aff34},\ref{aff35}}
\and S.~Casas\orcid{0000-0002-4751-5138}\inst{\ref{aff36}}
\and M.~Castellano\orcid{0000-0001-9875-8263}\inst{\ref{aff32}}
\and G.~Castignani\orcid{0000-0001-6831-0687}\inst{\ref{aff15}}
\and S.~Cavuoti\orcid{0000-0002-3787-4196}\inst{\ref{aff22},\ref{aff37}}
\and K.~C.~Chambers\orcid{0000-0001-6965-7789}\inst{\ref{aff38}}
\and A.~Cimatti\inst{\ref{aff39}}
\and C.~Colodro-Conde\inst{\ref{aff40}}
\and G.~Congedo\orcid{0000-0003-2508-0046}\inst{\ref{aff41}}
\and C.~J.~Conselice\orcid{0000-0003-1949-7638}\inst{\ref{aff42}}
\and L.~Conversi\orcid{0000-0002-6710-8476}\inst{\ref{aff43},\ref{aff44}}
\and Y.~Copin\orcid{0000-0002-5317-7518}\inst{\ref{aff45}}
\and A.~Costille\inst{\ref{aff46}}
\and F.~Courbin\orcid{0000-0003-0758-6510}\inst{\ref{aff47},\ref{aff48}}
\and H.~M.~Courtois\orcid{0000-0003-0509-1776}\inst{\ref{aff49}}
\and A.~Da~Silva\orcid{0000-0002-6385-1609}\inst{\ref{aff50},\ref{aff51}}
\and H.~Degaudenzi\orcid{0000-0002-5887-6799}\inst{\ref{aff52}}
\and S.~de~la~Torre\inst{\ref{aff46}}
\and G.~De~Lucia\orcid{0000-0002-6220-9104}\inst{\ref{aff10}}
\and F.~Dubath\orcid{0000-0002-6533-2810}\inst{\ref{aff52}}
\and C.~A.~J.~Duncan\orcid{0009-0003-3573-0791}\inst{\ref{aff41},\ref{aff42}}
\and X.~Dupac\inst{\ref{aff44}}
\and M.~Farina\orcid{0000-0002-3089-7846}\inst{\ref{aff53}}
\and R.~Farinelli\inst{\ref{aff15}}
\and F.~Faustini\orcid{0000-0001-6274-5145}\inst{\ref{aff32},\ref{aff54}}
\and S.~Ferriol\inst{\ref{aff45}}
\and F.~Finelli\orcid{0000-0002-6694-3269}\inst{\ref{aff15},\ref{aff55}}
\and N.~Fourmanoit\orcid{0009-0005-6816-6925}\inst{\ref{aff4}}
\and M.~Frailis\orcid{0000-0002-7400-2135}\inst{\ref{aff10}}
\and E.~Franceschi\orcid{0000-0002-0585-6591}\inst{\ref{aff15}}
\and M.~Fumana\orcid{0000-0001-6787-5950}\inst{\ref{aff8}}
\and S.~Galeotta\orcid{0000-0002-3748-5115}\inst{\ref{aff10}}
\and K.~George\orcid{0000-0002-1734-8455}\inst{\ref{aff56}}
\and B.~Gillis\orcid{0000-0002-4478-1270}\inst{\ref{aff41}}
\and C.~Giocoli\orcid{0000-0002-9590-7961}\inst{\ref{aff15},\ref{aff18}}
\and J.~Gracia-Carpio\inst{\ref{aff57}}
\and A.~Grazian\orcid{0000-0002-5688-0663}\inst{\ref{aff58}}
\and F.~Grupp\inst{\ref{aff57},\ref{aff56}}
\and L.~Guzzo\orcid{0000-0001-8264-5192}\inst{\ref{aff59},\ref{aff14},\ref{aff60}}
\and S.~V.~H.~Haugan\orcid{0000-0001-9648-7260}\inst{\ref{aff61}}
\and W.~Holmes\inst{\ref{aff62}}
\and F.~Hormuth\inst{\ref{aff63}}
\and A.~Hornstrup\orcid{0000-0002-3363-0936}\inst{\ref{aff64},\ref{aff65}}
\and K.~Jahnke\orcid{0000-0003-3804-2137}\inst{\ref{aff66}}
\and M.~Jhabvala\inst{\ref{aff67}}
\and B.~Joachimi\orcid{0000-0001-7494-1303}\inst{\ref{aff68}}
\and E.~Keih\"anen\orcid{0000-0003-1804-7715}\inst{\ref{aff69}}
\and S.~Kermiche\orcid{0000-0002-0302-5735}\inst{\ref{aff4}}
\and B.~Kubik\orcid{0009-0006-5823-4880}\inst{\ref{aff45}}
\and M.~Kunz\orcid{0000-0002-3052-7394}\inst{\ref{aff70}}
\and H.~Kurki-Suonio\orcid{0000-0002-4618-3063}\inst{\ref{aff71},\ref{aff72}}
\and A.~M.~C.~Le~Brun\orcid{0000-0002-0936-4594}\inst{\ref{aff7}}
\and S.~Ligori\orcid{0000-0003-4172-4606}\inst{\ref{aff28}}
\and P.~B.~Lilje\orcid{0000-0003-4324-7794}\inst{\ref{aff61}}
\and V.~Lindholm\orcid{0000-0003-2317-5471}\inst{\ref{aff71},\ref{aff72}}
\and I.~Lloro\orcid{0000-0001-5966-1434}\inst{\ref{aff73}}
\and G.~Mainetti\orcid{0000-0003-2384-2377}\inst{\ref{aff74}}
\and D.~Maino\inst{\ref{aff59},\ref{aff8},\ref{aff60}}
\and E.~Maiorano\orcid{0000-0003-2593-4355}\inst{\ref{aff15}}
\and O.~Mansutti\orcid{0000-0001-5758-4658}\inst{\ref{aff10}}
\and O.~Marggraf\orcid{0000-0001-7242-3852}\inst{\ref{aff75}}
\and K.~Markovic\orcid{0000-0001-6764-073X}\inst{\ref{aff62}}
\and M.~Martinelli\orcid{0000-0002-6943-7732}\inst{\ref{aff32},\ref{aff33}}
\and N.~Martinet\orcid{0000-0003-2786-7790}\inst{\ref{aff46}}
\and F.~Marulli\orcid{0000-0002-8850-0303}\inst{\ref{aff76},\ref{aff15},\ref{aff18}}
\and R.~Massey\orcid{0000-0002-6085-3780}\inst{\ref{aff77}}
\and E.~Medinaceli\orcid{0000-0002-4040-7783}\inst{\ref{aff15}}
\and Y.~Mellier\inst{\ref{aff78},\ref{aff9}}
\and M.~Meneghetti\orcid{0000-0003-1225-7084}\inst{\ref{aff15},\ref{aff18}}
\and E.~Merlin\orcid{0000-0001-6870-8900}\inst{\ref{aff32}}
\and G.~Meylan\inst{\ref{aff79}}
\and A.~Mora\orcid{0000-0002-1922-8529}\inst{\ref{aff80}}
\and M.~Moresco\orcid{0000-0002-7616-7136}\inst{\ref{aff76},\ref{aff15}}
\and L.~Moscardini\orcid{0000-0002-3473-6716}\inst{\ref{aff76},\ref{aff15},\ref{aff18}}
\and C.~Neissner\orcid{0000-0001-8524-4968}\inst{\ref{aff81},\ref{aff35}}
\and S.-M.~Niemi\orcid{0009-0005-0247-0086}\inst{\ref{aff29}}
\and C.~Padilla\orcid{0000-0001-7951-0166}\inst{\ref{aff81}}
\and S.~Paltani\orcid{0000-0002-8108-9179}\inst{\ref{aff52}}
\and F.~Pasian\orcid{0000-0002-4869-3227}\inst{\ref{aff10}}
\and K.~Pedersen\inst{\ref{aff82}}
\and W.~J.~Percival\orcid{0000-0002-0644-5727}\inst{\ref{aff83},\ref{aff84},\ref{aff85}}
\and V.~Pettorino\inst{\ref{aff29}}
\and S.~Pires\orcid{0000-0002-0249-2104}\inst{\ref{aff86}}
\and G.~Polenta\orcid{0000-0003-4067-9196}\inst{\ref{aff54}}
\and M.~Poncet\inst{\ref{aff87}}
\and L.~A.~Popa\inst{\ref{aff88}}
\and F.~Raison\orcid{0000-0002-7819-6918}\inst{\ref{aff57}}
\and A.~Renzi\orcid{0000-0001-9856-1970}\inst{\ref{aff89},\ref{aff90}}
\and J.~Rhodes\orcid{0000-0002-4485-8549}\inst{\ref{aff62}}
\and G.~Riccio\inst{\ref{aff22}}
\and F.~Rizzo\orcid{0000-0002-9407-585X}\inst{\ref{aff10}}
\and E.~Romelli\orcid{0000-0003-3069-9222}\inst{\ref{aff10}}
\and M.~Roncarelli\orcid{0000-0001-9587-7822}\inst{\ref{aff15}}
\and R.~Saglia\orcid{0000-0003-0378-7032}\inst{\ref{aff56},\ref{aff57}}
\and Z.~Sakr\orcid{0000-0002-4823-3757}\inst{\ref{aff91},\ref{aff92},\ref{aff93}}
\and A.~G.~S\'anchez\orcid{0000-0003-1198-831X}\inst{\ref{aff57}}
\and D.~Sapone\orcid{0000-0001-7089-4503}\inst{\ref{aff94}}
\and B.~Sartoris\orcid{0000-0003-1337-5269}\inst{\ref{aff56},\ref{aff10}}
\and P.~Schneider\orcid{0000-0001-8561-2679}\inst{\ref{aff75}}
\and T.~Schrabback\orcid{0000-0002-6987-7834}\inst{\ref{aff95},\ref{aff75}}
\and M.~Scodeggio\inst{\ref{aff8}}
\and A.~Secroun\orcid{0000-0003-0505-3710}\inst{\ref{aff4}}
\and G.~Seidel\orcid{0000-0003-2907-353X}\inst{\ref{aff66}}
\and M.~Seiffert\orcid{0000-0002-7536-9393}\inst{\ref{aff62}}
\and S.~Serrano\orcid{0000-0002-0211-2861}\inst{\ref{aff2},\ref{aff96},\ref{aff3}}
\and P.~Simon\inst{\ref{aff75}}
\and C.~Sirignano\orcid{0000-0002-0995-7146}\inst{\ref{aff89}}
\and G.~Sirri\orcid{0000-0003-2626-2853}\inst{\ref{aff18}}
\and L.~Stanco\orcid{0000-0002-9706-5104}\inst{\ref{aff90}}
\and J.~Steinwagner\orcid{0000-0001-7443-1047}\inst{\ref{aff57}}
\and P.~Tallada-Cresp\'{i}\orcid{0000-0002-1336-8328}\inst{\ref{aff34},\ref{aff35}}
\and A.~N.~Taylor\inst{\ref{aff41}}
\and I.~Tereno\orcid{0000-0002-4537-6218}\inst{\ref{aff50},\ref{aff97}}
\and N.~Tessore\orcid{0000-0002-9696-7931}\inst{\ref{aff68}}
\and S.~Toft\orcid{0000-0003-3631-7176}\inst{\ref{aff98},\ref{aff99}}
\and R.~Toledo-Moreo\orcid{0000-0002-2997-4859}\inst{\ref{aff100}}
\and F.~Torradeflot\orcid{0000-0003-1160-1517}\inst{\ref{aff35},\ref{aff34}}
\and I.~Tutusaus\orcid{0000-0002-3199-0399}\inst{\ref{aff92}}
\and L.~Valenziano\orcid{0000-0002-1170-0104}\inst{\ref{aff15},\ref{aff55}}
\and J.~Valiviita\orcid{0000-0001-6225-3693}\inst{\ref{aff71},\ref{aff72}}
\and T.~Vassallo\orcid{0000-0001-6512-6358}\inst{\ref{aff56},\ref{aff10}}
\and A.~Veropalumbo\orcid{0000-0003-2387-1194}\inst{\ref{aff14},\ref{aff20},\ref{aff19}}
\and Y.~Wang\orcid{0000-0002-4749-2984}\inst{\ref{aff101}}
\and J.~Weller\orcid{0000-0002-8282-2010}\inst{\ref{aff56},\ref{aff57}}
\and G.~Zamorani\orcid{0000-0002-2318-301X}\inst{\ref{aff15}}
\and E.~Zucca\orcid{0000-0002-5845-8132}\inst{\ref{aff15}}
\and M.~Ballardini\orcid{0000-0003-4481-3559}\inst{\ref{aff102},\ref{aff103},\ref{aff15}}
\and E.~Bozzo\orcid{0000-0002-8201-1525}\inst{\ref{aff52}}
\and C.~Burigana\orcid{0000-0002-3005-5796}\inst{\ref{aff104},\ref{aff55}}
\and R.~Cabanac\orcid{0000-0001-6679-2600}\inst{\ref{aff92}}
\and M.~Calabrese\orcid{0000-0002-2637-2422}\inst{\ref{aff105},\ref{aff8}}
\and D.~Di~Ferdinando\inst{\ref{aff18}}
\and J.~A.~Escartin~Vigo\inst{\ref{aff57}}
\and L.~Gabarra\orcid{0000-0002-8486-8856}\inst{\ref{aff106}}
\and J.~Mart\'{i}n-Fleitas\orcid{0000-0002-8594-569X}\inst{\ref{aff107}}
\and S.~Matthew\orcid{0000-0001-8448-1697}\inst{\ref{aff41}}
\and N.~Mauri\orcid{0000-0001-8196-1548}\inst{\ref{aff39},\ref{aff18}}
\and R.~B.~Metcalf\orcid{0000-0003-3167-2574}\inst{\ref{aff76},\ref{aff15}}
\and A.~Pezzotta\orcid{0000-0003-0726-2268}\inst{\ref{aff14}}
\and M.~P\"ontinen\orcid{0000-0001-5442-2530}\inst{\ref{aff71}}
\and C.~Porciani\orcid{0000-0002-7797-2508}\inst{\ref{aff75}}
\and I.~Risso\orcid{0000-0003-2525-7761}\inst{\ref{aff14},\ref{aff20}}
\and V.~Scottez\orcid{0009-0008-3864-940X}\inst{\ref{aff78},\ref{aff108}}
\and M.~Sereno\orcid{0000-0003-0302-0325}\inst{\ref{aff15},\ref{aff18}}
\and M.~Tenti\orcid{0000-0002-4254-5901}\inst{\ref{aff18}}
\and M.~Viel\orcid{0000-0002-2642-5707}\inst{\ref{aff11},\ref{aff10},\ref{aff16},\ref{aff12},\ref{aff109}}
\and M.~Wiesmann\orcid{0009-0000-8199-5860}\inst{\ref{aff61}}
\and Y.~Akrami\orcid{0000-0002-2407-7956}\inst{\ref{aff110},\ref{aff111}}
\and S.~Alvi\orcid{0000-0001-5779-8568}\inst{\ref{aff102}}
\and I.~T.~Andika\orcid{0000-0001-6102-9526}\inst{\ref{aff112},\ref{aff113}}
\and S.~Anselmi\orcid{0000-0002-3579-9583}\inst{\ref{aff90},\ref{aff89},\ref{aff114}}
\and M.~Archidiacono\orcid{0000-0003-4952-9012}\inst{\ref{aff59},\ref{aff60}}
\and F.~Atrio-Barandela\orcid{0000-0002-2130-2513}\inst{\ref{aff115}}
\and D.~Bertacca\orcid{0000-0002-2490-7139}\inst{\ref{aff89},\ref{aff58},\ref{aff90}}
\and M.~Bethermin\orcid{0000-0002-3915-2015}\inst{\ref{aff116}}
\and A.~Blanchard\orcid{0000-0001-8555-9003}\inst{\ref{aff92}}
\and S.~Borgani\orcid{0000-0001-6151-6439}\inst{\ref{aff117},\ref{aff11},\ref{aff10},\ref{aff12},\ref{aff109}}
\and M.~L.~Brown\orcid{0000-0002-0370-8077}\inst{\ref{aff42}}
\and S.~Bruton\orcid{0000-0002-6503-5218}\inst{\ref{aff118}}
\and A.~Calabro\orcid{0000-0003-2536-1614}\inst{\ref{aff32}}
\and B.~Camacho~Quevedo\orcid{0000-0002-8789-4232}\inst{\ref{aff11},\ref{aff16},\ref{aff10}}
\and F.~Caro\inst{\ref{aff32}}
\and C.~S.~Carvalho\inst{\ref{aff97}}
\and T.~Castro\orcid{0000-0002-6292-3228}\inst{\ref{aff10},\ref{aff12},\ref{aff11},\ref{aff109}}
\and F.~Cogato\orcid{0000-0003-4632-6113}\inst{\ref{aff76},\ref{aff15}}
\and S.~Conseil\orcid{0000-0002-3657-4191}\inst{\ref{aff45}}
\and S.~Contarini\orcid{0000-0002-9843-723X}\inst{\ref{aff57}}
\and A.~R.~Cooray\orcid{0000-0002-3892-0190}\inst{\ref{aff119}}
\and S.~Davini\orcid{0000-0003-3269-1718}\inst{\ref{aff20}}
\and G.~Desprez\orcid{0000-0001-8325-1742}\inst{\ref{aff120}}
\and A.~D\'iaz-S\'anchez\orcid{0000-0003-0748-4768}\inst{\ref{aff121}}
\and J.~J.~Diaz\orcid{0000-0003-2101-1078}\inst{\ref{aff40}}
\and S.~Di~Domizio\orcid{0000-0003-2863-5895}\inst{\ref{aff19},\ref{aff20}}
\and J.~M.~Diego\orcid{0000-0001-9065-3926}\inst{\ref{aff122}}
\and A.~Enia\orcid{0000-0002-0200-2857}\inst{\ref{aff17},\ref{aff15}}
\and Y.~Fang\inst{\ref{aff56}}
\and A.~G.~Ferrari\orcid{0009-0005-5266-4110}\inst{\ref{aff18}}
\and A.~Finoguenov\orcid{0000-0002-4606-5403}\inst{\ref{aff71}}
\and A.~Franco\orcid{0000-0002-4761-366X}\inst{\ref{aff123},\ref{aff124},\ref{aff125}}
\and K.~Ganga\orcid{0000-0001-8159-8208}\inst{\ref{aff126}}
\and J.~Garc\'ia-Bellido\orcid{0000-0002-9370-8360}\inst{\ref{aff110}}
\and T.~Gasparetto\orcid{0000-0002-7913-4866}\inst{\ref{aff32}}
\and V.~Gautard\inst{\ref{aff127}}
\and E.~Gaztanaga\orcid{0000-0001-9632-0815}\inst{\ref{aff3},\ref{aff2},\ref{aff128}}
\and F.~Giacomini\orcid{0000-0002-3129-2814}\inst{\ref{aff18}}
\and F.~Gianotti\orcid{0000-0003-4666-119X}\inst{\ref{aff15}}
\and G.~Gozaliasl\orcid{0000-0002-0236-919X}\inst{\ref{aff129},\ref{aff71}}
\and M.~Guidi\orcid{0000-0001-9408-1101}\inst{\ref{aff17},\ref{aff15}}
\and C.~M.~Gutierrez\orcid{0000-0001-7854-783X}\inst{\ref{aff130}}
\and A.~Hall\orcid{0000-0002-3139-8651}\inst{\ref{aff41}}
\and C.~Hern\'andez-Monteagudo\orcid{0000-0001-5471-9166}\inst{\ref{aff131},\ref{aff40}}
\and H.~Hildebrandt\orcid{0000-0002-9814-3338}\inst{\ref{aff132}}
\and J.~Hjorth\orcid{0000-0002-4571-2306}\inst{\ref{aff82}}
\and J.~J.~E.~Kajava\orcid{0000-0002-3010-8333}\inst{\ref{aff133},\ref{aff134}}
\and Y.~Kang\orcid{0009-0000-8588-7250}\inst{\ref{aff52}}
\and V.~Kansal\orcid{0000-0002-4008-6078}\inst{\ref{aff135},\ref{aff136}}
\and D.~Karagiannis\orcid{0000-0002-4927-0816}\inst{\ref{aff102},\ref{aff137}}
\and K.~Kiiveri\inst{\ref{aff69}}
\and C.~C.~Kirkpatrick\inst{\ref{aff69}}
\and S.~Kruk\orcid{0000-0001-8010-8879}\inst{\ref{aff44}}
\and M.~Lattanzi\orcid{0000-0003-1059-2532}\inst{\ref{aff103}}
\and J.~Le~Graet\orcid{0000-0001-6523-7971}\inst{\ref{aff4}}
\and L.~Legrand\orcid{0000-0003-0610-5252}\inst{\ref{aff138},\ref{aff139}}
\and M.~Lembo\orcid{0000-0002-5271-5070}\inst{\ref{aff9}}
\and F.~Lepori\orcid{0009-0000-5061-7138}\inst{\ref{aff140}}
\and G.~Leroy\orcid{0009-0004-2523-4425}\inst{\ref{aff141},\ref{aff77}}
\and G.~F.~Lesci\orcid{0000-0002-4607-2830}\inst{\ref{aff76},\ref{aff15}}
\and J.~Lesgourgues\orcid{0000-0001-7627-353X}\inst{\ref{aff36}}
\and L.~Leuzzi\orcid{0009-0006-4479-7017}\inst{\ref{aff15}}
\and T.~I.~Liaudat\orcid{0000-0002-9104-314X}\inst{\ref{aff142}}
\and J.~Macias-Perez\orcid{0000-0002-5385-2763}\inst{\ref{aff143}}
\and G.~Maggio\orcid{0000-0003-4020-4836}\inst{\ref{aff10}}
\and M.~Magliocchetti\orcid{0000-0001-9158-4838}\inst{\ref{aff53}}
\and F.~Mannucci\orcid{0000-0002-4803-2381}\inst{\ref{aff144}}
\and R.~Maoli\orcid{0000-0002-6065-3025}\inst{\ref{aff145},\ref{aff32}}
\and C.~J.~A.~P.~Martins\orcid{0000-0002-4886-9261}\inst{\ref{aff146},\ref{aff23}}
\and L.~Maurin\orcid{0000-0002-8406-0857}\inst{\ref{aff147}}
\and M.~Miluzio\inst{\ref{aff44},\ref{aff148}}
\and P.~Monaco\orcid{0000-0003-2083-7564}\inst{\ref{aff117},\ref{aff10},\ref{aff12},\ref{aff11}}
\and C.~Moretti\orcid{0000-0003-3314-8936}\inst{\ref{aff10},\ref{aff16},\ref{aff11},\ref{aff12}}
\and G.~Morgante\inst{\ref{aff15}}
\and S.~Nadathur\orcid{0000-0001-9070-3102}\inst{\ref{aff128}}
\and K.~Naidoo\orcid{0000-0002-9182-1802}\inst{\ref{aff128}}
\and A.~Navarro-Alsina\orcid{0000-0002-3173-2592}\inst{\ref{aff75}}
\and S.~Nesseris\orcid{0000-0002-0567-0324}\inst{\ref{aff110}}
\and L.~Pagano\orcid{0000-0003-1820-5998}\inst{\ref{aff102},\ref{aff103}}
\and F.~Passalacqua\orcid{0000-0002-8606-4093}\inst{\ref{aff89},\ref{aff90}}
\and K.~Paterson\orcid{0000-0001-8340-3486}\inst{\ref{aff66}}
\and L.~Patrizii\inst{\ref{aff18}}
\and A.~Pisani\orcid{0000-0002-6146-4437}\inst{\ref{aff4}}
\and D.~Potter\orcid{0000-0002-0757-5195}\inst{\ref{aff140}}
\and S.~Quai\orcid{0000-0002-0449-8163}\inst{\ref{aff76},\ref{aff15}}
\and M.~Radovich\orcid{0000-0002-3585-866X}\inst{\ref{aff58}}
\and P.~Reimberg\orcid{0000-0003-3410-0280}\inst{\ref{aff78}}
\and P.-F.~Rocci\inst{\ref{aff147}}
\and G.~Rodighiero\orcid{0000-0002-9415-2296}\inst{\ref{aff89},\ref{aff58}}
\and S.~Sacquegna\orcid{0000-0002-8433-6630}\inst{\ref{aff149},\ref{aff124},\ref{aff123}}
\and M.~Sahl\'en\orcid{0000-0003-0973-4804}\inst{\ref{aff150}}
\and D.~B.~Sanders\orcid{0000-0002-1233-9998}\inst{\ref{aff38}}
\and E.~Sarpa\orcid{0000-0002-1256-655X}\inst{\ref{aff16},\ref{aff109},\ref{aff12}}
\and A.~Schneider\orcid{0000-0001-7055-8104}\inst{\ref{aff140}}
\and D.~Sciotti\orcid{0009-0008-4519-2620}\inst{\ref{aff32},\ref{aff33}}
\and E.~Sellentin\inst{\ref{aff151},\ref{aff31}}
\and L.~C.~Smith\orcid{0000-0002-3259-2771}\inst{\ref{aff152}}
\and J.~G.~Sorce\orcid{0000-0002-2307-2432}\inst{\ref{aff153},\ref{aff147}}
\and K.~Tanidis\orcid{0000-0001-9843-5130}\inst{\ref{aff106}}
\and C.~Tao\orcid{0000-0001-7961-8177}\inst{\ref{aff4}}
\and G.~Testera\inst{\ref{aff20}}
\and R.~Teyssier\orcid{0000-0001-7689-0933}\inst{\ref{aff154}}
\and S.~Tosi\orcid{0000-0002-7275-9193}\inst{\ref{aff19},\ref{aff20},\ref{aff14}}
\and A.~Troja\orcid{0000-0003-0239-4595}\inst{\ref{aff89},\ref{aff90}}
\and M.~Tucci\inst{\ref{aff52}}
\and C.~Valieri\inst{\ref{aff18}}
\and A.~Venhola\orcid{0000-0001-6071-4564}\inst{\ref{aff155}}
\and D.~Vergani\orcid{0000-0003-0898-2216}\inst{\ref{aff15}}
\and F.~Vernizzi\orcid{0000-0003-3426-2802}\inst{\ref{aff156}}
\and G.~Verza\orcid{0000-0002-1886-8348}\inst{\ref{aff157}}
\and P.~Vielzeuf\orcid{0000-0003-2035-9339}\inst{\ref{aff4}}
\and N.~A.~Walton\orcid{0000-0003-3983-8778}\inst{\ref{aff152}}}
										   
\institute{Aix-Marseille Universit\'e, Universit\'e de Toulon, CNRS, CPT, Marseille, France\label{aff1}
\and
Institut d'Estudis Espacials de Catalunya (IEEC),  Edifici RDIT, Campus UPC, 08860 Castelldefels, Barcelona, Spain\label{aff2}
\and
Institute of Space Sciences (ICE, CSIC), Campus UAB, Carrer de Can Magrans, s/n, 08193 Barcelona, Spain\label{aff3}
\and
Aix-Marseille Universit\'e, CNRS/IN2P3, CPPM, Marseille, France\label{aff4}
\and
Aix Marseille Univ, INSERM, MMG, Marseille, France\label{aff5}
\and
Center for Data-Driven Discovery, Kavli IPMU (WPI), UTIAS, The University of Tokyo, Kashiwa, Chiba 277-8583, Japan\label{aff6}
\and
Laboratoire d'etude de l'Univers et des phenomenes eXtremes, Observatoire de Paris, Universit\'e PSL, Sorbonne Universit\'e, CNRS, 92190 Meudon, France\label{aff7}
\and
INAF-IASF Milano, Via Alfonso Corti 12, 20133 Milano, Italy\label{aff8}
\and
Institut d'Astrophysique de Paris, UMR 7095, CNRS, and Sorbonne Universit\'e, 98 bis boulevard Arago, 75014 Paris, France\label{aff9}
\and
INAF-Osservatorio Astronomico di Trieste, Via G. B. Tiepolo 11, 34143 Trieste, Italy\label{aff10}
\and
IFPU, Institute for Fundamental Physics of the Universe, via Beirut 2, 34151 Trieste, Italy\label{aff11}
\and
INFN, Sezione di Trieste, Via Valerio 2, 34127 Trieste TS, Italy\label{aff12}
\and
School of Mathematics and Physics, University of Surrey, Guildford, Surrey, GU2 7XH, UK\label{aff13}
\and
INAF-Osservatorio Astronomico di Brera, Via Brera 28, 20122 Milano, Italy\label{aff14}
\and
INAF-Osservatorio di Astrofisica e Scienza dello Spazio di Bologna, Via Piero Gobetti 93/3, 40129 Bologna, Italy\label{aff15}
\and
SISSA, International School for Advanced Studies, Via Bonomea 265, 34136 Trieste TS, Italy\label{aff16}
\and
Dipartimento di Fisica e Astronomia, Universit\`a di Bologna, Via Gobetti 93/2, 40129 Bologna, Italy\label{aff17}
\and
INFN-Sezione di Bologna, Viale Berti Pichat 6/2, 40127 Bologna, Italy\label{aff18}
\and
Dipartimento di Fisica, Universit\`a di Genova, Via Dodecaneso 33, 16146, Genova, Italy\label{aff19}
\and
INFN-Sezione di Genova, Via Dodecaneso 33, 16146, Genova, Italy\label{aff20}
\and
Department of Physics "E. Pancini", University Federico II, Via Cinthia 6, 80126, Napoli, Italy\label{aff21}
\and
INAF-Osservatorio Astronomico di Capodimonte, Via Moiariello 16, 80131 Napoli, Italy\label{aff22}
\and
Instituto de Astrof\'isica e Ci\^encias do Espa\c{c}o, Universidade do Porto, CAUP, Rua das Estrelas, PT4150-762 Porto, Portugal\label{aff23}
\and
Faculdade de Ci\^encias da Universidade do Porto, Rua do Campo de Alegre, 4150-007 Porto, Portugal\label{aff24}
\and
European Southern Observatory, Karl-Schwarzschild-Str.~2, 85748 Garching, Germany\label{aff25}
\and
Dipartimento di Fisica, Universit\`a degli Studi di Torino, Via P. Giuria 1, 10125 Torino, Italy\label{aff26}
\and
INFN-Sezione di Torino, Via P. Giuria 1, 10125 Torino, Italy\label{aff27}
\and
INAF-Osservatorio Astrofisico di Torino, Via Osservatorio 20, 10025 Pino Torinese (TO), Italy\label{aff28}
\and
European Space Agency/ESTEC, Keplerlaan 1, 2201 AZ Noordwijk, The Netherlands\label{aff29}
\and
Institute Lorentz, Leiden University, Niels Bohrweg 2, 2333 CA Leiden, The Netherlands\label{aff30}
\and
Leiden Observatory, Leiden University, Einsteinweg 55, 2333 CC Leiden, The Netherlands\label{aff31}
\and
INAF-Osservatorio Astronomico di Roma, Via Frascati 33, 00078 Monteporzio Catone, Italy\label{aff32}
\and
INFN-Sezione di Roma, Piazzale Aldo Moro, 2 - c/o Dipartimento di Fisica, Edificio G. Marconi, 00185 Roma, Italy\label{aff33}
\and
Centro de Investigaciones Energ\'eticas, Medioambientales y Tecnol\'ogicas (CIEMAT), Avenida Complutense 40, 28040 Madrid, Spain\label{aff34}
\and
Port d'Informaci\'{o} Cient\'{i}fica, Campus UAB, C. Albareda s/n, 08193 Bellaterra (Barcelona), Spain\label{aff35}
\and
Institute for Theoretical Particle Physics and Cosmology (TTK), RWTH Aachen University, 52056 Aachen, Germany\label{aff36}
\and
INFN section of Naples, Via Cinthia 6, 80126, Napoli, Italy\label{aff37}
\and
Institute for Astronomy, University of Hawaii, 2680 Woodlawn Drive, Honolulu, HI 96822, USA\label{aff38}
\and
Dipartimento di Fisica e Astronomia "Augusto Righi" - Alma Mater Studiorum Universit\`a di Bologna, Viale Berti Pichat 6/2, 40127 Bologna, Italy\label{aff39}
\and
Instituto de Astrof\'{\i}sica de Canarias, V\'{\i}a L\'actea, 38205 La Laguna, Tenerife, Spain\label{aff40}
\and
Institute for Astronomy, University of Edinburgh, Royal Observatory, Blackford Hill, Edinburgh EH9 3HJ, UK\label{aff41}
\and
Jodrell Bank Centre for Astrophysics, Department of Physics and Astronomy, University of Manchester, Oxford Road, Manchester M13 9PL, UK\label{aff42}
\and
European Space Agency/ESRIN, Largo Galileo Galilei 1, 00044 Frascati, Roma, Italy\label{aff43}
\and
ESAC/ESA, Camino Bajo del Castillo, s/n., Urb. Villafranca del Castillo, 28692 Villanueva de la Ca\~nada, Madrid, Spain\label{aff44}
\and
Universit\'e Claude Bernard Lyon 1, CNRS/IN2P3, IP2I Lyon, UMR 5822, Villeurbanne, F-69100, France\label{aff45}
\and
Aix-Marseille Universit\'e, CNRS, CNES, LAM, Marseille, France\label{aff46}
\and
Institut de Ci\`{e}ncies del Cosmos (ICCUB), Universitat de Barcelona (IEEC-UB), Mart\'{i} i Franqu\`{e}s 1, 08028 Barcelona, Spain\label{aff47}
\and
Instituci\'o Catalana de Recerca i Estudis Avan\c{c}ats (ICREA), Passeig de Llu\'{\i}s Companys 23, 08010 Barcelona, Spain\label{aff48}
\and
UCB Lyon 1, CNRS/IN2P3, IUF, IP2I Lyon, 4 rue Enrico Fermi, 69622 Villeurbanne, France\label{aff49}
\and
Departamento de F\'isica, Faculdade de Ci\^encias, Universidade de Lisboa, Edif\'icio C8, Campo Grande, PT1749-016 Lisboa, Portugal\label{aff50}
\and
Instituto de Astrof\'isica e Ci\^encias do Espa\c{c}o, Faculdade de Ci\^encias, Universidade de Lisboa, Campo Grande, 1749-016 Lisboa, Portugal\label{aff51}
\and
Department of Astronomy, University of Geneva, ch. d'Ecogia 16, 1290 Versoix, Switzerland\label{aff52}
\and
INAF-Istituto di Astrofisica e Planetologia Spaziali, via del Fosso del Cavaliere, 100, 00100 Roma, Italy\label{aff53}
\and
Space Science Data Center, Italian Space Agency, via del Politecnico snc, 00133 Roma, Italy\label{aff54}
\and
INFN-Bologna, Via Irnerio 46, 40126 Bologna, Italy\label{aff55}
\and
Universit\"ats-Sternwarte M\"unchen, Fakult\"at f\"ur Physik, Ludwig-Maximilians-Universit\"at M\"unchen, Scheinerstrasse 1, 81679 M\"unchen, Germany\label{aff56}
\and
Max Planck Institute for Extraterrestrial Physics, Giessenbachstr. 1, 85748 Garching, Germany\label{aff57}
\and
INAF-Osservatorio Astronomico di Padova, Via dell'Osservatorio 5, 35122 Padova, Italy\label{aff58}
\and
Dipartimento di Fisica "Aldo Pontremoli", Universit\`a degli Studi di Milano, Via Celoria 16, 20133 Milano, Italy\label{aff59}
\and
INFN-Sezione di Milano, Via Celoria 16, 20133 Milano, Italy\label{aff60}
\and
Institute of Theoretical Astrophysics, University of Oslo, P.O. Box 1029 Blindern, 0315 Oslo, Norway\label{aff61}
\and
Jet Propulsion Laboratory, California Institute of Technology, 4800 Oak Grove Drive, Pasadena, CA, 91109, USA\label{aff62}
\and
Felix Hormuth Engineering, Goethestr. 17, 69181 Leimen, Germany\label{aff63}
\and
Technical University of Denmark, Elektrovej 327, 2800 Kgs. Lyngby, Denmark\label{aff64}
\and
Cosmic Dawn Center (DAWN), Denmark\label{aff65}
\and
Max-Planck-Institut f\"ur Astronomie, K\"onigstuhl 17, 69117 Heidelberg, Germany\label{aff66}
\and
NASA Goddard Space Flight Center, Greenbelt, MD 20771, USA\label{aff67}
\and
Department of Physics and Astronomy, University College London, Gower Street, London WC1E 6BT, UK\label{aff68}
\and
Department of Physics and Helsinki Institute of Physics, Gustaf H\"allstr\"omin katu 2, 00014 University of Helsinki, Finland\label{aff69}
\and
Universit\'e de Gen\`eve, D\'epartement de Physique Th\'eorique and Centre for Astroparticle Physics, 24 quai Ernest-Ansermet, CH-1211 Gen\`eve 4, Switzerland\label{aff70}
\and
Department of Physics, P.O. Box 64, 00014 University of Helsinki, Finland\label{aff71}
\and
Helsinki Institute of Physics, Gustaf H{\"a}llstr{\"o}min katu 2, University of Helsinki, Helsinki, Finland\label{aff72}
\and
SKA Observatory, Jodrell Bank, Lower Withington, Macclesfield, Cheshire SK11 9FT, UK\label{aff73}
\and
Centre de Calcul de l'IN2P3/CNRS, 21 avenue Pierre de Coubertin 69627 Villeurbanne Cedex, France\label{aff74}
\and
Universit\"at Bonn, Argelander-Institut f\"ur Astronomie, Auf dem H\"ugel 71, 53121 Bonn, Germany\label{aff75}
\and
Dipartimento di Fisica e Astronomia "Augusto Righi" - Alma Mater Studiorum Universit\`a di Bologna, via Piero Gobetti 93/2, 40129 Bologna, Italy\label{aff76}
\and
Department of Physics, Institute for Computational Cosmology, Durham University, South Road, Durham, DH1 3LE, UK\label{aff77}
\and
Institut d'Astrophysique de Paris, 98bis Boulevard Arago, 75014, Paris, France\label{aff78}
\and
Institute of Physics, Laboratory of Astrophysics, Ecole Polytechnique F\'ed\'erale de Lausanne (EPFL), Observatoire de Sauverny, 1290 Versoix, Switzerland\label{aff79}
\and
Telespazio UK S.L. for European Space Agency (ESA), Camino bajo del Castillo, s/n, Urbanizacion Villafranca del Castillo, Villanueva de la Ca\~nada, 28692 Madrid, Spain\label{aff80}
\and
Institut de F\'{i}sica d'Altes Energies (IFAE), The Barcelona Institute of Science and Technology, Campus UAB, 08193 Bellaterra (Barcelona), Spain\label{aff81}
\and
DARK, Niels Bohr Institute, University of Copenhagen, Jagtvej 155, 2200 Copenhagen, Denmark\label{aff82}
\and
Waterloo Centre for Astrophysics, University of Waterloo, Waterloo, Ontario N2L 3G1, Canada\label{aff83}
\and
Department of Physics and Astronomy, University of Waterloo, Waterloo, Ontario N2L 3G1, Canada\label{aff84}
\and
Perimeter Institute for Theoretical Physics, Waterloo, Ontario N2L 2Y5, Canada\label{aff85}
\and
Universit\'e Paris-Saclay, Universit\'e Paris Cit\'e, CEA, CNRS, AIM, 91191, Gif-sur-Yvette, France\label{aff86}
\and
Centre National d'Etudes Spatiales -- Centre spatial de Toulouse, 18 avenue Edouard Belin, 31401 Toulouse Cedex 9, France\label{aff87}
\and
Institute of Space Science, Str. Atomistilor, nr. 409 M\u{a}gurele, Ilfov, 077125, Romania\label{aff88}
\and
Dipartimento di Fisica e Astronomia "G. Galilei", Universit\`a di Padova, Via Marzolo 8, 35131 Padova, Italy\label{aff89}
\and
INFN-Padova, Via Marzolo 8, 35131 Padova, Italy\label{aff90}
\and
Institut f\"ur Theoretische Physik, University of Heidelberg, Philosophenweg 16, 69120 Heidelberg, Germany\label{aff91}
\and
Institut de Recherche en Astrophysique et Plan\'etologie (IRAP), Universit\'e de Toulouse, CNRS, UPS, CNES, 14 Av. Edouard Belin, 31400 Toulouse, France\label{aff92}
\and
Universit\'e St Joseph; Faculty of Sciences, Beirut, Lebanon\label{aff93}
\and
Departamento de F\'isica, FCFM, Universidad de Chile, Blanco Encalada 2008, Santiago, Chile\label{aff94}
\and
Universit\"at Innsbruck, Institut f\"ur Astro- und Teilchenphysik, Technikerstr. 25/8, 6020 Innsbruck, Austria\label{aff95}
\and
Satlantis, University Science Park, Sede Bld 48940, Leioa-Bilbao, Spain\label{aff96}
\and
Instituto de Astrof\'isica e Ci\^encias do Espa\c{c}o, Faculdade de Ci\^encias, Universidade de Lisboa, Tapada da Ajuda, 1349-018 Lisboa, Portugal\label{aff97}
\and
Cosmic Dawn Center (DAWN)\label{aff98}
\and
Niels Bohr Institute, University of Copenhagen, Jagtvej 128, 2200 Copenhagen, Denmark\label{aff99}
\and
Universidad Polit\'ecnica de Cartagena, Departamento de Electr\'onica y Tecnolog\'ia de Computadoras,  Plaza del Hospital 1, 30202 Cartagena, Spain\label{aff100}
\and
Infrared Processing and Analysis Center, California Institute of Technology, Pasadena, CA 91125, USA\label{aff101}
\and
Dipartimento di Fisica e Scienze della Terra, Universit\`a degli Studi di Ferrara, Via Giuseppe Saragat 1, 44122 Ferrara, Italy\label{aff102}
\and
Istituto Nazionale di Fisica Nucleare, Sezione di Ferrara, Via Giuseppe Saragat 1, 44122 Ferrara, Italy\label{aff103}
\and
INAF, Istituto di Radioastronomia, Via Piero Gobetti 101, 40129 Bologna, Italy\label{aff104}
\and
Astronomical Observatory of the Autonomous Region of the Aosta Valley (OAVdA), Loc. Lignan 39, I-11020, Nus (Aosta Valley), Italy\label{aff105}
\and
Department of Physics, Oxford University, Keble Road, Oxford OX1 3RH, UK\label{aff106}
\and
Aurora Technology for European Space Agency (ESA), Camino bajo del Castillo, s/n, Urbanizacion Villafranca del Castillo, Villanueva de la Ca\~nada, 28692 Madrid, Spain\label{aff107}
\and
ICL, Junia, Universit\'e Catholique de Lille, LITL, 59000 Lille, France\label{aff108}
\and
ICSC - Centro Nazionale di Ricerca in High Performance Computing, Big Data e Quantum Computing, Via Magnanelli 2, Bologna, Italy\label{aff109}
\and
Instituto de F\'isica Te\'orica UAM-CSIC, Campus de Cantoblanco, 28049 Madrid, Spain\label{aff110}
\and
CERCA/ISO, Department of Physics, Case Western Reserve University, 10900 Euclid Avenue, Cleveland, OH 44106, USA\label{aff111}
\and
Technical University of Munich, TUM School of Natural Sciences, Physics Department, James-Franck-Str.~1, 85748 Garching, Germany\label{aff112}
\and
Max-Planck-Institut f\"ur Astrophysik, Karl-Schwarzschild-Str.~1, 85748 Garching, Germany\label{aff113}
\and
Laboratoire Univers et Th\'eorie, Observatoire de Paris, Universit\'e PSL, Universit\'e Paris Cit\'e, CNRS, 92190 Meudon, France\label{aff114}
\and
Departamento de F{\'\i}sica Fundamental. Universidad de Salamanca. Plaza de la Merced s/n. 37008 Salamanca, Spain\label{aff115}
\and
Universit\'e de Strasbourg, CNRS, Observatoire astronomique de Strasbourg, UMR 7550, 67000 Strasbourg, France\label{aff116}
\and
Dipartimento di Fisica - Sezione di Astronomia, Universit\`a di Trieste, Via Tiepolo 11, 34131 Trieste, Italy\label{aff117}
\and
California Institute of Technology, 1200 E California Blvd, Pasadena, CA 91125, USA\label{aff118}
\and
Department of Physics \& Astronomy, University of California Irvine, Irvine CA 92697, USA\label{aff119}
\and
Kapteyn Astronomical Institute, University of Groningen, PO Box 800, 9700 AV Groningen, The Netherlands\label{aff120}
\and
Departamento F\'isica Aplicada, Universidad Polit\'ecnica de Cartagena, Campus Muralla del Mar, 30202 Cartagena, Murcia, Spain\label{aff121}
\and
Instituto de F\'isica de Cantabria, Edificio Juan Jord\'a, Avenida de los Castros, 39005 Santander, Spain\label{aff122}
\and
INFN, Sezione di Lecce, Via per Arnesano, CP-193, 73100, Lecce, Italy\label{aff123}
\and
Department of Mathematics and Physics E. De Giorgi, University of Salento, Via per Arnesano, CP-I93, 73100, Lecce, Italy\label{aff124}
\and
INAF-Sezione di Lecce, c/o Dipartimento Matematica e Fisica, Via per Arnesano, 73100, Lecce, Italy\label{aff125}
\and
Universit\'e Paris Cit\'e, CNRS, Astroparticule et Cosmologie, 75013 Paris, France\label{aff126}
\and
CEA Saclay, DFR/IRFU, Service d'Astrophysique, Bat. 709, 91191 Gif-sur-Yvette, France\label{aff127}
\and
Institute of Cosmology and Gravitation, University of Portsmouth, Portsmouth PO1 3FX, UK\label{aff128}
\and
Department of Computer Science, Aalto University, PO Box 15400, Espoo, FI-00 076, Finland\label{aff129}
\and
Instituto de Astrof\'\i sica de Canarias, c/ Via Lactea s/n, La Laguna 38200, Spain. Departamento de Astrof\'\i sica de la Universidad de La Laguna, Avda. Francisco Sanchez, La Laguna, 38200, Spain\label{aff130}
\and
Universidad de La Laguna, Departamento de Astrof\'{\i}sica, 38206 La Laguna, Tenerife, Spain\label{aff131}
\and
Ruhr University Bochum, Faculty of Physics and Astronomy, Astronomical Institute (AIRUB), German Centre for Cosmological Lensing (GCCL), 44780 Bochum, Germany\label{aff132}
\and
Department of Physics and Astronomy, Vesilinnantie 5, 20014 University of Turku, Finland\label{aff133}
\and
Serco for European Space Agency (ESA), Camino bajo del Castillo, s/n, Urbanizacion Villafranca del Castillo, Villanueva de la Ca\~nada, 28692 Madrid, Spain\label{aff134}
\and
ARC Centre of Excellence for Dark Matter Particle Physics, Melbourne, Australia\label{aff135}
\and
Centre for Astrophysics \& Supercomputing, Swinburne University of Technology,  Hawthorn, Victoria 3122, Australia\label{aff136}
\and
Department of Physics and Astronomy, University of the Western Cape, Bellville, Cape Town, 7535, South Africa\label{aff137}
\and
DAMTP, Centre for Mathematical Sciences, Wilberforce Road, Cambridge CB3 0WA, UK\label{aff138}
\and
Kavli Institute for Cosmology Cambridge, Madingley Road, Cambridge, CB3 0HA, UK\label{aff139}
\and
Department of Astrophysics, University of Zurich, Winterthurerstrasse 190, 8057 Zurich, Switzerland\label{aff140}
\and
Department of Physics, Centre for Extragalactic Astronomy, Durham University, South Road, Durham, DH1 3LE, UK\label{aff141}
\and
IRFU, CEA, Universit\'e Paris-Saclay 91191 Gif-sur-Yvette Cedex, France\label{aff142}
\and
Univ. Grenoble Alpes, CNRS, Grenoble INP, LPSC-IN2P3, 53, Avenue des Martyrs, 38000, Grenoble, France\label{aff143}
\and
INAF-Osservatorio Astrofisico di Arcetri, Largo E. Fermi 5, 50125, Firenze, Italy\label{aff144}
\and
Dipartimento di Fisica, Sapienza Universit\`a di Roma, Piazzale Aldo Moro 2, 00185 Roma, Italy\label{aff145}
\and
Centro de Astrof\'{\i}sica da Universidade do Porto, Rua das Estrelas, 4150-762 Porto, Portugal\label{aff146}
\and
Universit\'e Paris-Saclay, CNRS, Institut d'astrophysique spatiale, 91405, Orsay, France\label{aff147}
\and
HE Space for European Space Agency (ESA), Camino bajo del Castillo, s/n, Urbanizacion Villafranca del Castillo, Villanueva de la Ca\~nada, 28692 Madrid, Spain\label{aff148}
\and
INAF - Osservatorio Astronomico d'Abruzzo, Via Maggini, 64100, Teramo, Italy\label{aff149}
\and
Theoretical astrophysics, Department of Physics and Astronomy, Uppsala University, Box 516, 751 37 Uppsala, Sweden\label{aff150}
\and
Mathematical Institute, University of Leiden, Einsteinweg 55, 2333 CA Leiden, The Netherlands\label{aff151}
\and
Institute of Astronomy, University of Cambridge, Madingley Road, Cambridge CB3 0HA, UK\label{aff152}
\and
Univ. Lille, CNRS, Centrale Lille, UMR 9189 CRIStAL, 59000 Lille, France\label{aff153}
\and
Department of Astrophysical Sciences, Peyton Hall, Princeton University, Princeton, NJ 08544, USA\label{aff154}
\and
Space physics and astronomy research unit, University of Oulu, Pentti Kaiteran katu 1, FI-90014 Oulu, Finland\label{aff155}
\and
Institut de Physique Th\'eorique, CEA, CNRS, Universit\'e Paris-Saclay 91191 Gif-sur-Yvette Cedex, France\label{aff156}
\and
Center for Computational Astrophysics, Flatiron Institute, 162 5th Avenue, 10010, New York, NY, USA\label{aff157}}    

 \title{\Euclid\/ preparation}
 
 \subtitle{LXXXVII. Non-Gaussianity of two-point statistics likelihood: Precise analysis of the matter power spectrum distribution}

\abstract{

We investigate the non-Gaussian features in the distribution of the matter power spectrum multipoles. Using the \covmos~method, we generated 100\,000 mock realisations of dark matter density fields in both real and redshift space across multiple redshifts and cosmological models. We derived an analytical framework linking the non-Gaussianity of the power spectrum distribution to higher-order statistics of the density field, including the trispectrum and pentaspectrum. We explored the effect of redshift-space distortions, the geometry of the survey, the Fourier binning, the integral constraint, and
the shot noise on the skewness of the distribution of the power spectrum measurements. Our results demonstrate that the likelihood of the estimated matter power spectrum significantly deviates from a Gaussian assumption on non-linear scales, particularly at low redshift. This departure is primarily driven by the pentaspectrum contribution, which dominates over the trispectrum at intermediate scales. We also examined the impact of the finiteness of the survey geometry in the context of the \Euclid mission, and we find that both the shape of the survey and the integral constraint amplify the skewness. 

}
\keywords{Non-Gaussian Likelihood -- Euclid -- Large-scale structures -- 2-point statistics}

   \titlerunning{Non-Gaussianity of two-point statistics likelihood}
   \authorrunning{ Euclid Collaboration: J.~Bel et al.}
   
   \maketitle
%
%
%
%
   
\section{Introduction}

In recent years, observational cosmology has experienced significant advancements with the emergence of next-generation galaxy surveys designed to probe larger cosmic volumes and smaller scales than ever before. Examples of such surveys include the \Euclid mission \citep{EuclidSkyOverview}, the Vera C. Rubin Observatory Legacy Survey of Space and Time \citep[LSST,][]{ivezic2018lsst}, and the Dark Energy Spectroscopic Instrument  \citep[DESI,][]{aghamousa_16}. These cutting-edge surveys will provide unprecedented insights into the structure and evolution of the Universe, making it increasingly important to develop more accurate and robust methodologies for data analysis. As we push the boundaries of observational capabilities, it is crucial that our methods of parameter inference can ensure that we can effectively extract valuable information from the wealth of data that will be gathered by these surveys.

Cosmological information is generally extracted from the analysis of two-point statistics of the fields under study. In real space, this is done with the two-point correlation function, and in Fourier space, it is done with the power spectrum. The latter is the case for state-of-the art analyses of the temperature and polarisation of the cosmic microwave background \citep[CMB;][]{planck_20} or the large-scale structure (LSS) of the Universe, using galaxy clustering obtained through spectroscopic data \citep{gilmarin_20, bautista_21, DESI_24} or a combination of weak lensing and clustering  obtained through photometric data \citep{DES3x2_22, KiDS_21, HSC_23}. 

To extract this cosmological information, we need to assume, a priory, what the distribution of our data is in order  to model the likelihood function. A simple assumption that is commonly made for two-point statistics is that they follow a multivariate Gaussian distribution. However, we know that it is not the case, even for Gaussian fields such as the CMB. Indeed, the two-point statistics of a Gaussian field follows a $\chi^2$ distribution with a number of degrees of freedom equal to the number of independent modes at the considered scale. On large scales this number is low such that the $\chi^2$ distribution is far from being a Gaussian \citep{hamimeche_08}.

Moreover, at intermediate scales (above $30\ \Mpc$), such as those probed by the \Euclid mission \citep{EuclidSkyOverview}, the non-linear clustering of matter generates non-vanishing $N$-point statistics (with $N>2$) of the density field.
A well-known consequence of the non-Gaussianity of the density field from the non-linear clustering of matter on the two-point statistics is the non-Gaussian contribution to the covariance matrix from a non-vanishing four-point correlation function \citep{scoccimarro_99}. This higher-order contribution to the covariance is still present even though the likelihood is approximated to be Gaussian. 
In addition, these higher-order correlation functions also induce higher-order moments in the distribution of the two-point statistics, thus causing departure from the Gaussian assumption also at small scales.
Indeed, the skewness and kurtosis of the two-point statistics distribution have been shown to be non-vanishing on small scales and especially at low redshift  \citep{takahashi_09, blot_15, lin_20, upham_21}. \citet{hall_22} also studied the potential influence of higher-order $N$-point statistics on the non-Gaussian distribution of two-point statistics on large scales in the context of a weak lensing likelihood analysis, but they found it to be negligible.

Given the precision with which we aim to measure the cosmological parameters with \Euclid, we need to understand and control in detail the likelihood modelling and test whether it can significantly bias our cosmological analyses. Indeed, cosmological constraints derived from likelihood analysis generally rely on the assumption that the measurements are distributed according to a Gaussian. As a result, if the Gaussian assumption is not justified, it might affect the derived cosmological parameters. This is the aim of the present paper and its companion, Gouyou Beauchamps et al. (in prep., hereafter cited as GB-2025).
In this work, we focus particularly on the Fourier version of two-point statistics, the power spectrum, to develop an analytical framework to understand how the non-Gaussianity of the two-point statistics distribution is linked to the non-Gaussianity of the field itself and what are the settings of the likelihood to which it is the most sensitive. In particular, we look at how the skewness and kurtosis of the distribution of the matter power spectrum changes with respect to (i) cosmological effects such as the cosmological model, the redshift, and the presence or absence of redshift space distortions (RSD); (ii) the settings of the power spectrum estimator, such as the presence of aliasing and the Fourier mode binning; and (iii) the effects originating from specific features of the survey, such as the sample density and the survey footprint. Our companion paper, GB-2025, specifically focuses on the impact of wrongly assuming a Gaussian likelihood on \Euclid cosmological constraints.

To achieve all of this, we extensively rely on the \covmos~method \citep{baratta_19, baratta_22} in order to produce very large sets of approximate simulations of non-Gaussian fields, which allow us to precisely explore the distribution of the power spectrum. It has been shown in \citet{baratta_22} that \covmos~is a reliable tool to model the covariance of two-point statistics, and we show in this paper that it is also the case for the study of their full distribution. 

Since the main scope of the present paper is to understand the mechanisms possibly producing a non-Gaussian behaviour of the distribution of the estimated power spectrum, we focus on the matter field. In GB-2025, we use more realistic mock galaxy catalogues to study the skewness of the galaxy power spectrum. 
Finally, we note that this work does not aim to study the non-Gaussianity of the likelihood arising from the estimation of a covariance matrix with a low number of realisations \citep{Sellentin16, Percival22}.

The paper is structured as follows. In Sect. \ref{sect:theory} we present our analytical framework to study the distribution of the power spectrum. In Sect. \ref{sect:simu} we describe the simulation sets we constructed with \covmos~and introduce the estimators we applied on them. In Sect. \ref{sect:measure} we study the estimated distribution of the power spectrum under various scenarios. In Sect. \ref{sect:single_mode} we study the distribution of individual Fourier modes. Finally, we conclude in Sect. \ref{sect:conclu}.

\section{Skewness of the power spectrum estimator}
\label{sect:theory}
In this section we derive the skewness of the probability distribution of the estimator of the power spectrum in terms of the higher-order $N$-points cumulant moments of the density field.
We start with some definitions. The ensemble average of the random variable $X$ following the probability density function (PDF) $\pcal$ is $\langle X \rangle$, and $\langle X \rangle_{\rm c} $ represents its connected expectation value. In practice, the random variable we study is the estimator of the multipoles of the power spectrum, $\hat P^{(\ell)}(k)$, defined  as

\begin{equation}
\hat P^{(\ell)}(k) := k_{\rm F}^3\,\frac{2\ell+1}{N_k}\sum_{i=1}^{N_k} |\delta_{\vec k_i}|^2\, \lcal_\ell(\mu_i)\;,
\label{estimatorpk}
\end{equation}
where $\ell$ refers to the order of the considered multipole, $\lcal_\ell$ are the Legendre polynomials of order $\ell$, $k$ is the modulus of the considered spherical $k$-shell in Fourier space, $N_k$ is the number of independent modes that it contains, $\mu_i$ is the cosine of the angle between the line of sight and the wave vector $\vec k_i$ and the sum is made over independent modes inside the $k$-shell. Note that, by independent we do not mean statistically independent, but we rather refer to the fact that since the configuration space density field $\delta(\vec x)$ is real, its Fourier space counterpart $\delta_{\vec k}$ satisfies $\delta_{-\vec k} = \delta_{\vec k}^\star$, where the $\star$ stands for complex conjugate. This means that there are repeated modes, that are not considered in Eq.~\eqref{estimatorpk}. Unless otherwise stated, we consider a periodic boundary\footnote{We use the following Fourier convention $\delta_{\vec k}:=\!\frac{1}{(2\pi)^3}\int_{L^3}\!\delta(\vec x) \, {\rm e}^{-{\rm i}\vec k\cdot \vec x}\dif^3\vec x.$}{conditions} of size $L$, thus characterised by its fundamental mode $k_{\rm F}:= 2\pi/L$. That is the reason why the sum in Eq.~\eqref{estimatorpk} is discrete. Finally, due to statistical invariance by translation (i.e. statistical homogeneity) one can define the power spectrum ($P$), the bispectrum ($B$), the trispectrum ($T$), and the pentaspectrum ($Q$) as

\begin{eqnarray}
k_{\rm F}^3\, \langle \delta_{\vec k_1}\delta_{\vec k_2}\rangle_{\rm{c}} & = & \delta_{\vec k_{12}}^\mathrm{K} P(\vec k_1)\; , \\
k_{\rm F}^3\, \langle \delta_{\vec k_1}\delta_{\vec k_2}\delta_{\vec k_3}\rangle_{\rm c}  & = & \delta_{\vec k_{123}}^\mathrm{K} B(\vec k_1, \vec k_2)\; , \label{bispec} \\
k_{\rm F}^3\, \langle \delta_{\vec k_1}\delta_{\vec k_2}\delta_{\vec k_3}\delta_{\vec k_4}\rangle_{\rm c}  & = & \delta_{\vec k_{1234}}^\mathrm{K} T(\vec k_1, \vec k_2, \vec k_3)\; , \\
k_{\rm F}^3\, \langle \delta_{\vec k_1}\delta_{\vec k_2}\delta_{\vec k_3}\delta_{\vec k_4}\delta_{\vec k_5}\delta_{\vec k_6}\rangle_{\rm c}  & = & \delta_{\vec k_{123456}}^\mathrm{K} Q(\vec k_1, \vec k_2, \vec k_3,\vec k_4, \vec k_5)\; ,
\end{eqnarray}
where we use the specific notation $\delta_{\vec k_{1..N}}^\mathrm{K}$ for the Kronecker symbol, which is $1$ if $\vec k_1 + ... + \vec k_N = \vec 0$ and $0$ otherwise. To simplify the following calculations, we defined the random variables $X:= \hat P^{(\ell)}(k)$, $X_i:= |\delta_{\vec k_i}|^2$ and the coefficients $a_i:= k_{\rm F}^3\frac{2\ell+1}{N_k}\lcal_\ell(\mu_i)$. Thus, $X_i$ depends on the wave vector $\vec k$, and $X$ depends on the wave mode $k$ and on the chosen multipole $\ell$.
Based on these definitions, Eq.~\eqref{estimatorpk} takes the form
\begin{equation}
X = \sum_{i=1}^{N_k} a_i\, X_i \;.
\label{estimatorx}
\end{equation}

Since the multi-point probability distribution function $\pcal( X_1, ...\, , X_{N_k} )$ of the modes $X_i$ is not known, one can only focus on the cumulant moments in order to characterise how far from a Gaussian the distribution of $X$ is. Indeed, if the variable $X$ were following a Gaussian distribution then its cumulant moments should be $\langle X^n\rangle_{\rm c} =0$ for $n \geq 3$. Taking the ensemble average of Eq.~\eqref{estimatorx} one can show that the mean of $X$ is given by 

\begin{equation}
\begin{array}{rcl}
\langle X\rangle & = & \frac{2\ell+1}{N_k}\sum_{i=1}^{N_k}P(\vec k_i)\, \lcal_\ell(\mu_i) \\
& \simeq &\frac{2\ell+1}{2}\int_{-1}^{1}P(\vec k)\, \lcal_\ell(\mu)\, \dif \mu = P^{(\ell)}(k)\;,
\end{array}
\label{meanb}
\end{equation}
which means that the estimator is unbiased in the large $N_k$ limit.
We can then show that its variance is given by 

\begin{equation}
\langle X^2\rangle_{\rm c}  = \sum_{i=1}^{N_k} a_i^2\,\langle X_i^2\rangle_{\rm c}  + 2 \sum_{j>i}a_i a_j\, \langle X_i X_j\rangle_{\rm c} \;, 
\label{var}
\end{equation}
where the second sum is made over the $N_k(N_k-1)/2$ permutations of $i$ and $j$ without repetitions. If the Fourier modes were statistically independent inside the shell then one could drop the double sum (second term) in Eq.~\eqref{var}. Notice that there is a subtlety that needs to be mentioned when inspecting Eq.~\eqref{var}, despite the fact that we define $X_i:=|\delta_{\vec k_i}|^2$ one cannot say that $\langle X_i^n\rangle_{\rm c}  = \langle |\delta_{\vec k_i}|^{2n} \rangle_{\rm c} $. The equality holds only for the moments but not for the cumulant moments. Indeed, without paying attention to this, looking at Eq.~\eqref{var} one would erroneously conclude that for a Gaussian field the variance of the shell average vanishes. Instead, one can use the relation between moments $\langle X_i^n\rangle = \langle |\delta_{\vec k_i}|^{2n} \rangle$ together with the assumption of statistical invariance by translation to show that for one-point statistics

\begin{equation}
\langle X_i^2 \rangle_{\rm c}  = \langle |\delta_{\vec k_i}|^4 \rangle_{\rm c}  + \langle |\delta_{\vec k_i}|^2 \rangle_{\rm c} ^2\;, 
\label{varb}
\end{equation}
while for two-point statistics 

\begin{equation}
\langle X_i X_j \rangle_{\rm c}  = \langle |\delta_{\vec k_i}|^2 |\delta_{\vec k_j}|^2\rangle_{\rm c} \;.
\label{crossb}
\end{equation}
As a result, Eq.~\eqref{var}  $\langle X_iX_j\rangle_{\rm c} $ involves a four-point function of the density field that is null in the Gaussian case. Instead, the first term of Eq.~\eqref{var} contains an intrinsic contribution that would arise even if the density field were Gaussian (i.e. $\langle |\delta_{\vec k}|^2\rangle_{\rm c} ^2$) and a part that would only arise when considering a non-Gaussian density field (i.e. $\langle |\delta_{\vec k}|^4\rangle_{\rm c} $). The usual \citep[see][]{scoccimarro_99, meiksin_99, Smith_2009, Smith_2015} expression for the variance of the power spectrum can be obtained by splitting $\langle X_i^2 \rangle_{\rm c} $ (see Eq.~\ref{varb}) into its Gaussian and non-Gaussian parts:

\begin{equation}
\langle X^2\rangle_{\rm c}  = \frac{\bar P_\ell^2}{N_k} + k_{\rm F}^3 \bar T_\ell\;,  
\label{usualvar}
\end{equation}
where we define $k$ as being at the centre of the spherical shell of thickness $k_{\rm F}$ and $V_k$ its volume. In addition, we define the shell average powers of the power spectrum as 

\begin{equation}
\bar P_\ell^n := \frac{(2\ell+1)^n}{V_k}\int_{V_k} P^n(\vec k_1)\, \lcal_\ell^n(\mu_1)\, \dif^3\vec k_1\;,
\label{pnbar}
\end{equation}
where the subscript $V_k$ in the integral means that the integration variables belong to the Fourier shell, $k$, and thus $\bar P_\ell^n$ depends on the Fourier shell $k$. We note that we used the continuous limit of the shell average as described in the appendix \ref{appendixa}. We also defined the corresponding quantity for the trispectrum as

\begin{equation}
\bar T_\ell := \frac{(2\ell+1)^2}{V_k^2}\int_{V_k} T(\vec k_1, -\vec k_1, \vec k_2)\, \lcal_\ell(\mu_1)\lcal_\ell(\mu_2)\, \dif^3\vec k_1\dif^3\vec k_2\;,
\label{tnbar}
\end{equation}
which also depends on the considered $k$-shell.

The same reasoning about moments and cumulant moments can be applied to obtain the formal expression of the skewness: 

\begin{equation}
\begin{array}{rl}
\langle X^3\rangle_{\rm c}  = & \displaystyle \sum_{i=1}^{N_k} a_i^3\langle X_i^3\rangle_{\rm c}  + 3 \sum_{j>i}\left [ a_i^2 a_j \langle X_i^2 X_j\rangle_{\rm c}  + a_i a_j^2 \langle X_i X_j^2\rangle_{\rm c} \right ] \\
 & \\
 & + 6 \displaystyle  \sum_{n>j>i} a_ia_ja_n \langle X_i X_j X_n\rangle_{\rm c} \;  . \label{skew}
\end{array}
\end{equation}
Equation~\eqref{skew} allows us to understand that three kinds of correlators can contribute to the skewness, and they are 

\begin{eqnarray}
\langle X_i^3 \rangle_{\rm c}  & = &\langle |\delta_{\vec k_i}|^6 \rangle_{\rm c}  + 6 \langle |\delta_{\vec k_i}|^4 \rangle_{\rm c} \langle |\delta_{\vec k_i}|^2 \rangle_{\rm c}  + 2 \langle |\delta_{\vec k_i}|^2 \rangle_{\rm c} ^3\;, \label{termone}\\
\langle X_i^2 X_j \rangle_{\rm c}  & = &  \langle |\delta_{\vec k_i}|^4 |\delta_{\vec k_j}|^2\rangle_{\rm c}  + 2 \langle |\delta_{\vec k_i}|^2 \rangle_{\rm c}  \langle |\delta_{\vec k_i}|^2 |\delta_{\vec k_j}|^2\rangle_{\rm c}\;,   \label{termtwo} \\
\langle X_i X_j X_n\rangle_{\rm c}   & = &\langle |\delta_{\vec k_i}|^2 |\delta_{\vec k_j}|^2|\delta_{\vec k_n}|^2\rangle_{\rm c}  + \langle \delta_{\vec k_i}\delta_{\vec k_j}\delta_{\vec k_n}\rangle_{\rm c} ^2  \nonumber \\
& & \!\!\!\!\! + \langle \delta_{-\vec k_i}\delta_{\vec k_j}\delta_{\vec k_n}\rangle_{\rm c} ^2 + \!\langle \delta_{\vec k_i}\delta_{-\vec k_j}\delta_{\vec k_n}\rangle_{\rm c} ^2  + \!\langle \delta_{\vec k_i}\delta_{\vec k_j}\delta_{-\vec k_n}\rangle_{\rm c} ^2\, . \label{tripoint}
\end{eqnarray}

The one-point cumulant (Eq.~\ref{termone}) is composed of an intrinsic contribution (present even in the Gaussian case) and contributions related to the pentaspectrum and trispectrum. Then, there are two-point correlators (Eq.~\ref{termtwo}) and three-point correlators (Eq.~\ref{tripoint}) which would both be null in the Gaussian case since they involve only six- and four-point functions. By inserting Eqs.\,\eqref{termone}--\eqref{tripoint}, 
into Eq.\,\eqref{skew}, one can express the full expression of the skewness as

\begin{equation}
\langle X^3\rangle_{\rm c}  = \frac{2}{N_k^2} \bar P_\ell^3 + \frac{3}{N_k} k_{\rm F}^3 \left ( \bar B_\ell^2 + 2\bar M_\ell \right ) + k_{\rm F}^6 \bar Q_\ell\;,
\label{skewnessfull}
\end{equation}
where we define the shell average of the square of the bispectrum as

\begin{equation}
\begin{array}{rl}
\bar B_\ell^2 := \displaystyle \frac{(2\ell+1)^3}{V_k^2} & \displaystyle \int_{V_k} B^2(\vec k_1, \vec k_2)\, \lcal_\ell(\mu_1) \lcal_\ell(\mu_2) \\
& \\
& \times \lcal_\ell(\mu_1+\mu_2)\, \Theta_k(k_3)\, \dif^3\vec k_1\dif^3 \vec k_2\;;
\end{array}
\label{bl}
\end{equation}
the shell average of the product between the power spectrum and the trispectrum as

\begin{equation}
\begin{array}{rl}
\bar M_\ell := \dfrac{(2\ell+1)^3}{V_k^2} & \displaystyle\int_{V_k} P(\vec k_1)\, T(\vec k_1, -\vec k_1, \vec k_2)\\
&\\
&\times \lcal_\ell^2(\mu_1) \lcal_\ell(\mu_2)\, \dif^3\vec k_1\dif^3 \vec k_2\;;
\end{array}
\label{ml}
\end{equation}
and the shell averaged pentaspectrum as

\begin{equation}
\begin{array}{rl}
\bar{Q}_\ell := \displaystyle \frac{(2\ell+1)^3}{V_k^3} & \displaystyle \int_{V_k} Q(\vec k_1, -\vec k_1, \vec k_2, -\vec k_2, \vec k_3)\\ 
& \\
& \times \lcal_\ell(\mu_1) \lcal_\ell(\mu_2)\lcal_\ell(\mu_3)\,\dif^3\vec k_1\dif^3 \vec k_2 \dif^3 \vec k_3\;.
\label{ql}
\end{array}
\end{equation}

In the present work we aim at quantifying the various terms involved in Eq.~\eqref{skewnessfull} and their relative contribution to the reduced skewness, 
\begin{equation}
S_3^{(\ell)} := \langle X^3 \rangle_{\rm c} / \langle X^2\rangle_{\rm c}^{3/2},
\end{equation}
in various configurations of the density field. In the rest of the paper we refer to the reduced skewness, $S_3^{(\ell)}$, as the skewness, where $\ell$ refers to the fact that this is the skewness of the multipole $\ell$ of the power spectrum,\footnote{Note that this is an abuse of notation because $S_3^{(\ell)}$ is not actually a multipole of the Legendre expansion as $P^{(\ell)}(k)$.} and we refer to $\langle X^3 \rangle_{\rm c} $ as the third-order cumulant.

\section{Simulations and estimators}
\label{sect:simu}
In this section we first present the different sets of simulations we generated and analysed in this study. Then, we present the different estimators employed to probe the distribution of the estimated power spectra.

\subsection{Approximate simulations with \texttt{COVMOS}}
The standard $N$-body approach is a widely used technique for simulating the evolution of large-scale cosmic structures. In our case, however, the computational cost of producing the large number of realisations required for a robust analysis of the power spectrum distribution is prohibitive. 
This is why in this study we exclusively rely on an approximate mock-catalogue generator, \texttt{COVMOS}\footnote{  \href{https://github.com/PhilippeBaratta/COVMOS}{github.com/PhilippeBaratta/COVMOS}} \citep{baratta_19, baratta_22}.

The \texttt{COVMOS} public code enables rapid simulation of catalogues for various cosmological objects in both real and redshift space. Its speed is due to the fact that it does not require the evolution of an initial density, as typically needed in an $N$-body process. Instead, a density field is directly generated at a specific redshift and then subjected to local Poisson sampling to create the point-like catalogue.

Although similar approaches, such as log-normal mock generators, are extensively studied in the literature \citep{Xavier_16, Alonso_14, Agrawal_17}, the \texttt{COVMOS} method has two distinctive features. First, the power spectrum of the density and velocity fields, as well as the density one-point PDF, can be arbitrarily set by the user. This flexibility means it is not restricted to a log-normal PDF, enabling the exploration of more realistic models. These statistics can be provided either from analytical models or directly from estimates based on simulations or observational data.
    
The second distinctive feature of the \texttt{COVMOS} method is its ability to assign specific velocities to objects based on the local density, while ensuring the validity of the continuity equation. This capability leads to a high simulation performance for RSD effects on two-point statistics, covering both the linear and mildly non-linear regimes.

The \covmos\ method relies on the application of a local transform to a Gaussian field in order to match a target one-point PDF. As discussed in \citet{baratta_22}, this means that the higher order one-point moments of the PDF are correctly reproduced. Since they are given by specific integrals of the Fourier space $N$-point correlation functions, it is expected that \covmos\ is able to capture some of the higher order information. Indeed, in \citet{baratta_22} it has been shown that the non-Gaussian part of the covariance can be well recovered at least up to $k \sim 0.25h/$Mpc at $z=0$. In addition, as shown in Sect.~\ref{sect:theory}, the skewness of the distribution of the power spectrum depends on higher order correlation functions. This is the reason why we designed a validation procedure in Appendix \ref{validationofcovmos} that justifies the use of the \covmos\ method for this study.

In this study, we generated numerous large samples of dark matter catalogues in both real and redshift-space, all within simulated periodic boxes of volume $1\ (\mathrm{Gpc}/h)^3$. However, a cut in geometry of these catalogues, involving survey masks, is performed in Sect.~\ref{sect:mask}. These samples are detailed in Table~\ref{tablesamples}, which includes the sample name, simulated cosmology, and redshifts. Each sample contains 100\,000 realisations of $10^8$ particles, thus a density of 0.1 $(\mathrm{Mpc}/h)^{-3}$, this means that the shot-noise power spectrum is equal to the input power spectrum at $k=1.65\ \invMpc$.

We selected three different flat cosmologies compatible with the Planck 2013 results \citep{Planck_13}, with cosmological parameters $(\Omega_{\rm m}, \Omega_{\rm b}, h, n_{\rm s}, A_{\rm s} = (0.32, 0.05, 0.67, 0.96, 2.1265\times 10^{-9})$. The first cosmology is characterised by a Lambda cold dark matter ($\Lambda$CDM) background, the second (16nu) includes massive neutrinos with total mass of $\sum m_\nu = 0.16\ \mathrm{eV}$. In the third cosmology, named w9p3, a dynamical dark energy is included using the CPL \citep{chevallier_polarski_01, linder_03} parametrisation, where $w(a)= w_0 + w_a(1-a)$ and $(w_0,w_a)= (-0.9,0.3)$.

As previously stated, to use the \texttt{COVMOS} method, the user must specify target values for the power spectrum and one-point PDF of the density field, along with the target velocity power spectrum. For all our samples, we adopted these target values based on measurements performed in the \texttt{DEMNUni} suite of $N$-body simulations \citep{carbone_16} which have been run for the cosmological models described above.
In the case of the $\Lambda$CDM and 16nu cosmology we use the one-point PDF and power spectrum, averaged over the 50 \texttt{DEMNUni-Cov} realisations \citep{baratta_22, gouyou_25, parimbelli_21}, with a sampling of $512$ points per side. For the w9p3 cosmology we use the unique realisation of this cosmological model from the \texttt{DEMNUni} set of simulations \citep{parimbelli_22} with a sampling of $1024$ points per side (comoving box twice larger).

\begin{table}

\centering
\caption{Redshift at which the \texttt{COVMOS} mock catalogue samples considered in this work have been generated.}
\begin{tabular}{|l|c|c|c|c|c|}
  \hline
  name & $z$ \\
  \hline
   $\Lambda$CDM & $0,\, 0.5,\, 1,\, 1.5,\, 2$ \\
   16nu & $0$ \\
   w9p3 & $0$ \\
  \hline
\end{tabular}
\label{tablesamples}
\end{table}

\subsection{Statistics estimators}

To estimate the power spectrum, we employed the widely used \texttt{NBodyKit}\footnote{\href{https://nbodykit.readthedocs.io/en/latest/}{nbodykit.readthedocs.io}} software \citep{nbodykit}, which offers robust and efficient tools for analysing the multipoles of the power spectrum, both in real and redshift space. In particular it allows the user to vary different settings of the estimation procedure, including the type of mass assignment scheme employed to interpolate particles on a grid, or the activation or not of the interlacing technique allowing the user to reduce the aliasing effect. We take advantage of this opportunity to identify a possible impact of these settings on the distribution of the output power spectra.

Despite having introduced (in Sect. \ref{sect:theory}) only the lowest level of non-Gaussianity (i.e. the skewness) of the distribution of the measured power spectrum, in the present work we also estimate the reduced kurtosis $S_4^{(\ell)} := \langle X^4\rangle_{\rm c} /\langle X^2\rangle_{\rm c} ^2$. We apply the same nomenclature to the kurtosis, i.e. $S_4^{(\ell)}$ is dubbed kurtosis, and 
$\langle X^4\rangle_{\rm c} $ is dubbed the fourth-order cumulant.

The estimation of the skewness and kurtosis is based on the $N_{\rm s} = 100\,000$ realisations that we generated with the \covmos~method, and their estimators are
\begin{equation}
    \hat S_3^{(\ell)}(k) =   \frac{N_{\rm s}^{-1}\sum_{i=1}^{N_{\rm s}} \left[\hat P_i^{(\ell)} (k) - \bar P^{(\ell)}(k)\right]^3}{\left\{ N_{\rm s}^{-1}\sum_{i=1}^{N_{\rm s}} \left[\hat P_i^{(\ell)} (k) - \bar P^{(\ell)}(k)\right]^2 \right\}^{3/2}}\;
\label{s3estimator}
\end{equation}
and
\begin{equation}
    \hat S_4^{(\ell)}(k) =  \frac{N_{\rm s}^{-1}\sum_{i=1}^{N_{\rm s}} \left[\hat P_i^{(\ell)} (k) - \bar P^{(\ell)}(k)\right]^4}{\left\{ N_{\rm s}^{-1}\sum_{i=1}^{N_{\rm s}} \left[\hat P_i^{(\ell)} (k) - \bar P^{(\ell)}(k)\right]^2 \right\}^2} -3 \;,
    \label{s4estimator}
\end{equation}
where $\hat P_i^{(\ell)}(k)$ is our estimate of the $\ell$-th multipole of the power spectrum in shell $k$ for the $i$-th \covmos~realisation, and $\bar P^{(\ell)}(k)$ is the estimated average multipole power spectrum over the $N_{\rm s}$ \covmos~realisations, 

\begin{equation}
\bar P^{(\ell)}(k) := \frac{1}{N_{\rm s}} \sum_{i=1}^{N_{\rm s}} \hat P_i^{(\ell)} (k)\;.
\label{mean}
\end{equation}

Note that since we are estimating the mean $\bar P^{(\ell)}(k)$, the estimators of the skewness and kurtosis are expected to be biased. However, given the extremely high number of realisations, $N_{\rm s}$, we can safely neglect that bias. Indeed, the variance would be affected at the level of $N_{\rm s}/(N_{\rm s}-1)$, and the skewness would be affected at the level of $N_{\rm s}^2/(N_{\rm s}-2)/(N_{\rm s}-1)$ (i.e. adjusted Fisher--Pearson standardised moment coefficient), which is at most of the order of $0.3$\%.

The estimation of the skewness and kurtosis should be compared to the corresponding prediction assuming a Gaussian density field to quantify their excess. In Appendix~\ref{appendixa} we review the general Gaussian field prediction for $S_n^{(\ell)}$, which if one further assumes an isotropic real space power spectrum, it can be expressed as 

\begin{eqnarray}
S_{3, \rm{G}}^{(0)}(k) & = & \frac{2}{\sqrt{N_k}}\;, \label{s30gauss}\\
S_{4, \rm{G}}^{(0)}(k) & = & \frac{6}{N_k} \;,  \label{s40gauss}
\end{eqnarray}
for the monopole power spectrum estimator, where $N_k$ is the number of independent wave modes in the considered $k$-shell. One can notice that Eqs.~\eqref{s30gauss} and \eqref{s40gauss} apparently differ from \citet{takahashi_09}, but this is only due to a different definition of the number of modes, $N_k$. Indeed, they defined $N_k$ as the total number of modes, while we restrict it to the independent modes (as previously explained).  Thus, $N_k$ can either be computed exactly on a Fourier grid or approximately evaluated analytically by dividing the total Fourier volume of the shell ($2\pi\, k^2\, k_{\rm F}$) by the Fourier volume ($k_{\rm F}^3$) of an individual mode, yielding 

\begin{equation}
N_k \simeq 2\pi\, (k/k_{\rm F})^2\;.
\label{nkth}
\end{equation}
Note that Eqs.~\eqref{s30gauss} and \eqref{s40gauss} are different when considering the quadrupole and the hexadecapole. We show in Appendix~\ref{appendixa} that

\begin{eqnarray}
S_{3, \rm{G}}^{(2)}(k) & = & \sqrt{5}\frac{2}{7}\frac{2}{\sqrt{N_k}}\;, \\
S_{4, \rm{G}}^{(2)}(k) & = & \frac{15}{7}\frac{6}{N_k} \;,\\
S_{3, \rm{G}}^{(4)}(k) & = & 3\frac{162}{1001}\frac{2}{\sqrt{N_k}}\;, \\
S_{4, \rm{G}}^{(4)}(k) & = & 2.5180114\frac{6}{N_k}\;. \label{s44gauss}
\end{eqnarray}
In redshift space the expressions are more complicated but knowing the multipoles of the power spectrum one can still evaluate the corresponding expressions numerically on the 3D Fourier grid as given by Eq.~\eqref{cumulgauss} in Appendix~\ref{appendixa}.

\section{Quantifying the non-Gaussianity of the distribution of the estimated power spectrum}\label{sect:measure}

In this section we present a phenomenological study of the power spectrum distribution when varying different parameters. First, we study the cosmological effects that may impact the various statistics due to non-linear clustering. This includes the cosmological model and the redshift. Then, we investigate the effect of the power spectrum estimator's settings through the Fourier mode binning and the \texttt{NBodyKit} input parameters for the 3D Fourier grid. Finally, we look at survey specific features, namely the sample density and the survey mask. 

\subsection{Cosmological effects}

In this subsection we want to explore the potential cosmological factors that could influence the distribution of the power spectrum estimator, primarily as a result of non-linear clustering. Thanks to the analytical framework developed in Sect.~\ref{sect:theory} we are able to quantify the different terms contributing to the non-Gaussianity of the distribution.

\begin{figure*}
\includegraphics[width=\linewidth]{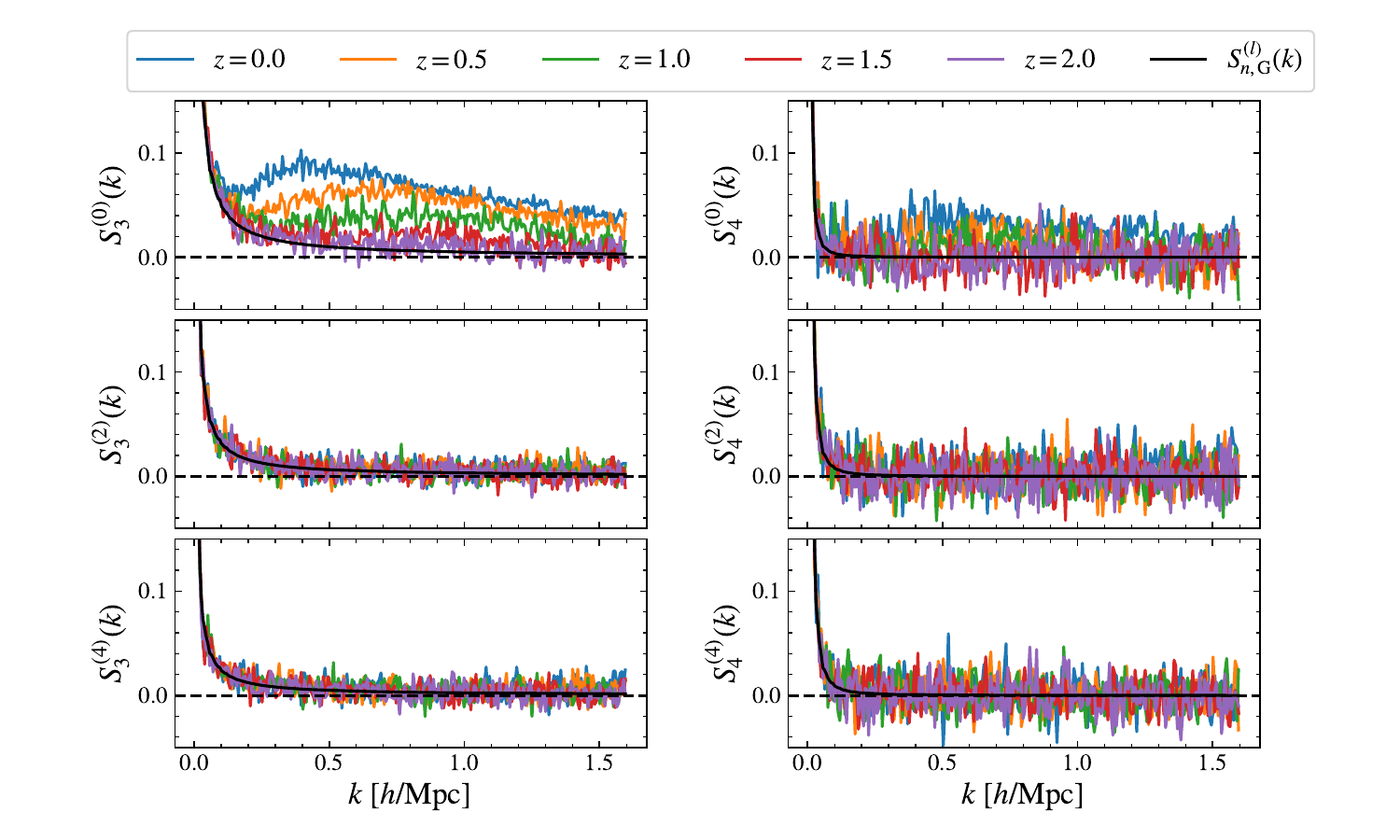}
\caption{Estimated skewness, $S_3^{(\ell)}$ (left panels), and kurtosis, $S_4^{(\ell)}$ (right panels), of the distribution of power spectrum multipoles $P^{(\ell)}(k)$  in real space for the reference $\Lambda$CDM cosmology and for the five redshifts considered in this work (from top to bottom panels $\ell = 0,2,4$). In each panel, the black line shows the prediction for a Gaussian field.}
\label{fig:S3S4multipolesreal}
\end{figure*}

\begin{figure*}
\includegraphics[width=\linewidth]{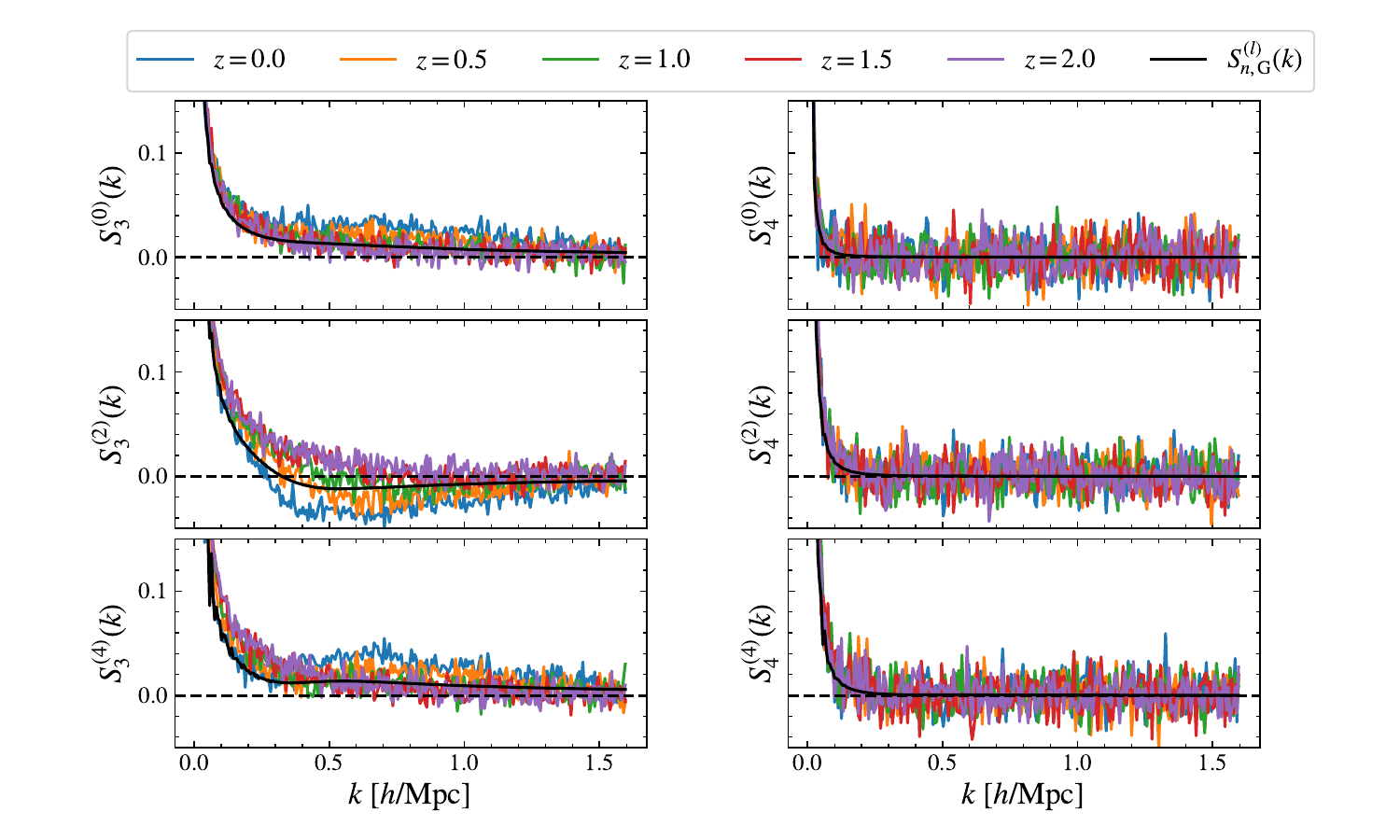}
\caption{Same as Fig.\ref{fig:S3S4multipolesreal} but in redshift space.}
\label{fig:S3S4multipolesredshiftspace}
\end{figure*}

In Fig.~\ref{fig:S3S4multipolesreal} we present the estimated skewness $S_3^{(\ell)}$ and kurtosis $S_4^{(\ell)}$ of the distribution of the multipoles of the power spectrum obtained from the sets of $\Lambda$CDM simulations at five different redshifts ranging from $z=0$ to $z=2$ in real space. The fact that we do not include redshift space distortions allowed us to compare these measurements to the Gaussian predictions of the skewness $S_3^{(\ell)}$ and kurtosis $S_4^{(\ell)}$ (see Eqs.\, \ref{s30gauss}--\ref{s44gauss}). One can see that on large scales (low $k$), both the skewness and kurtosis of the distributions are
systematically higher than zero, which agrees with the expectation for a Gaussian field (see Eqs.\, \ref{s30gauss}--\ref{s44gauss}). Indeed, as the wave number $k$ decreases, the number of wave modes, $N_k$, reduces, thus significantly increasing the skewness and kurtosis. 

Inspecting the left panel of Fig.~\ref{fig:S3S4multipolesreal}, we can see that only the high-$k$ part of the monopole's skewness shows significant deviation with respect to the Gaussian field prediction. This is in good agreement with \citet{blot_15} and with \citet{takahashi_09}. In addition, we can see that this deviation is more important at a low redshift. At $z=0$ we can see the excess of skewness starting at $k = 0.1\ \invMpc$, while at $z=1$ it starts at $k = 0.2\ \invMpc$, and we cannot even see it at $z=2$ for the scales probed here. This behaviour is expected as the terms $\bar B$, $\bar T$, and $\bar Q$ in Eq.~\eqref{skewnessfull} are becoming more important through the non-linear evolution of the matter clustering. Note that when $k$ increases, the skewness starts to decrease again but less rapidly than in the Gaussian case; thus one could think that in the limit of an infinite number of modes the skewness is expected to vanish, which is in agreement with the central limit theorem. 

Regarding the kurtosis (right panel of Fig.~\ref{fig:S3S4multipolesreal}) it is more difficult to draw strong conclusions because the estimation is more affected by noise. Focusing on the monopole we can detect some excess of kurtosis around $k>0.3\ \invMpc$ at $z=0$. However, for $z>0.5$ there is no direct detection of an excess kurtosis.

In redshift space (see Fig.~\ref{fig:S3S4multipolesredshiftspace}), the excess of skewness observed at high wave modes for the monopole in real space is suppressed, yielding a better agreement with the Gaussian prediction. The excess skewness of the monopole in real space appears to be transferred to the quadrupole and hexadecapole in redshift space, causing deviations from their respective predictions. Specifically, the skewness of the quadrupole decreases and becomes negative, while the skewness of the hexadecapole increases. Regarding the kurtosis, no excess can be detected at any scale, redshift, and multipole. 

Since it appears that the dominant non-Gaussian contribution is the skewness, in the following we put the kurtosis aside and quantify the contribution of each term in Eq.~\eqref{skewnessfull} to the total skewness. 

We define the relative difference of the total cumulant moments with respect to their Gaussian only contribution as
\be
r_i := \frac{\langle X^i\rangle_{\rm c} }{\langle X^i\rangle_{\rm c} ^{\rm G}}-1\;.
\ee
From Eq.~\eqref{usualvar}, one can show that $r_2$, which is the non-Gaussian relative contribution to the variance, is given by 
\begin{equation}
r_2 = N_k\, k_{\rm F}^3\, \frac{\bar T_0}{\bar P_0^2}\;.
\label{r2}
\end{equation}
It is directly related to the shell averaged trispectrum $\bar T_0$.

Based on the structure of Eq.~\eqref{skewnessfull}, we can write $r_3$ as

\begin{equation}
r_3 = \frac{3}{2} N_k k_{\rm F}^3\, \frac{\bar B_0^2 + 2\bar M_0}{\bar P_0^3} + \frac{N_k^2 k_{\rm F}^6}{2} \frac{\bar Q_0}{\bar P_0^3}\;.
\label{r3}
\end{equation}
When considering the monopole in real space, one can show that $\bar M_0 = P_0 \bar T_0$ and $\bar P_0^n = P_0^n$, which allows us to rewrite $r_3$ as 

\be\label{q0}
r_3 = b + 3r_2 + \frac{N_k^2 k_{\rm F}^6}{2} \frac{\bar Q_0}{P_0^3}\;,
\ee
where
\begin{equation}
b := \frac{3}{2}N_k k_{\rm F}^3 \bar B_0^2\;.
\label{bdef}
\end{equation}
Equation~\eqref{q0} highlights the fact that different terms contribute to the non-Gaussian part of $\langle X^3 \rangle_{\rm c}$, specifically: the bispectrun term $b$, the trispectrum term $3r_2$, and the pentaspectrum term.

\begin{figure}
\includegraphics[width=\linewidth]{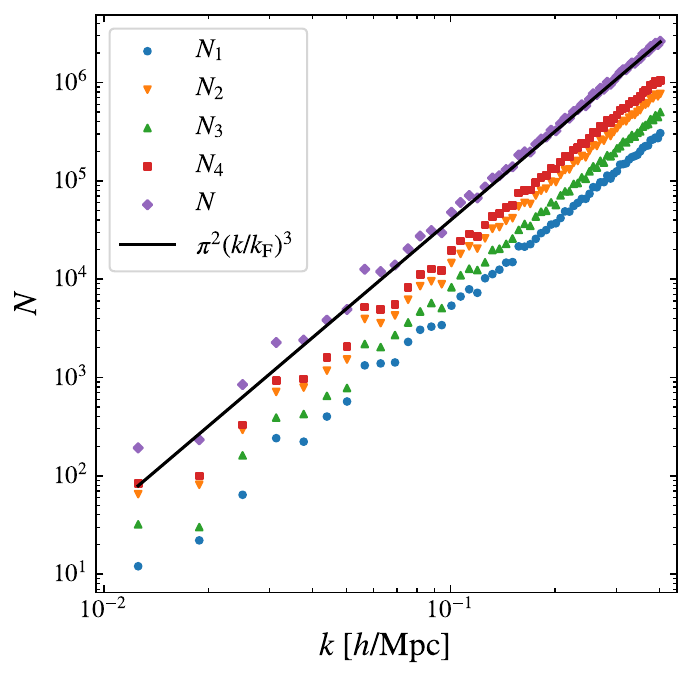}
\caption{Number of triplets, $\vec k_1$, $\vec k_2$, $\vec k_3$, per $k$-shell depending on the considered configuration, $N_1$, $N_2$, $N_3$, and $N_4$. The purple diamonds show the total number of triplets, and the black line shows the corresponding expected number.}
\label{fig:triplets}
\end{figure}

We start by showing that the bispectrum contribution, $b$, is expected to be negligible with respect to $3r_2$. For this we can notice two things about the three modes $\vec k_1$, $\vec k_2$, and $\vec k_3$ (see Eq. \ref{bispec}) defining a bispectrum configuration: first, invariance by translation imposes that $-\vec k_3= \vec k_1 + \vec k_2$, and second, we need their modulus to be contained within the considered $k$-shell. These two constraints mean that not all triplets, $\vec k_1$, $\vec k_2$, $\vec k_3$, will contribute to the shell average of the bispectrum. Indeed, only pairs $\vec k_1$, $\vec k_2$ that satisfy $|\vec k_1+\vec k_2|\simeq k$ can contribute. Thus, only nearly equilateral configurations of the triplets will be kept in the shell averaging. We can see in Eq.~\eqref{tripoint} that there are four kinds of configurations for the bispectrum $\vec k_1+\vec k_2$, $\vec k_1 - \vec k_2$, $-\vec k_1-\vec k_2$ and $\vec k_2 -\vec k_1$ with respectively $N_1$, $N_2$, $N_3$ and $N_4$ number of triangular configurations which are contributing to the shell-averaged square of the bispectrum $\bar B_0^2$. As a result, defining $N=N_1+N_2+N_3+N_4$ as the total number of triangles that contribute to the shell average and using symmetry conditions one can show that the total (i.e. double counting the repeated conjugate modes) number of triangles is roughly given by $N \simeq \pi^2\, (k/k_{\rm F})^3$. In addition, from Eq.~\eqref{bl} one can show that the monopole $\bar B_0$ is related to the equilateral configuration $B(k)$ of the bispectrum through
\be
\bar B_0^2 \simeq \frac{N}{N_k^2} B^2(k)\;.
\ee
In practice, when considering only half of the Fourier space (discarding non-independent modes) the four configurations of the triangles formed by $\vec k_1$ and $\vec k_2$ are not appearing with the same rate. This can be estimated directly on a Fourier grid and is represented in Fig.~\ref{fig:triplets}. In this figure we show that the total number $N$ of non-degenerate triangular configurations is indeed following the expected trend. One can immediately understand that since $N \propto k^3$ and $N_k^2 \propto k^4$ (see Eq.~\ref{nkth}) the ratio will lead to a suppression of the equilateral configuration of the bispectrum proportional to $k^{-1}$. We thus expect that the term $b$ is counting less than the trispectrum contribution, which is not suffering from this mode selection within the shell. In fact, assuming that $B(k) \simeq P^3(k)$ and that $\bar T(k) \simeq P^4(k)$ \citep[see][for a justification]{scoccimarro_99,baratta_19}, one can show that we expect

\begin{equation}
b\simeq 3 r_2\, \dfrac{N}{N_k^2}\;.
\label{aproxr3}
\end{equation}

\begin{figure}

\includegraphics[width=\linewidth]{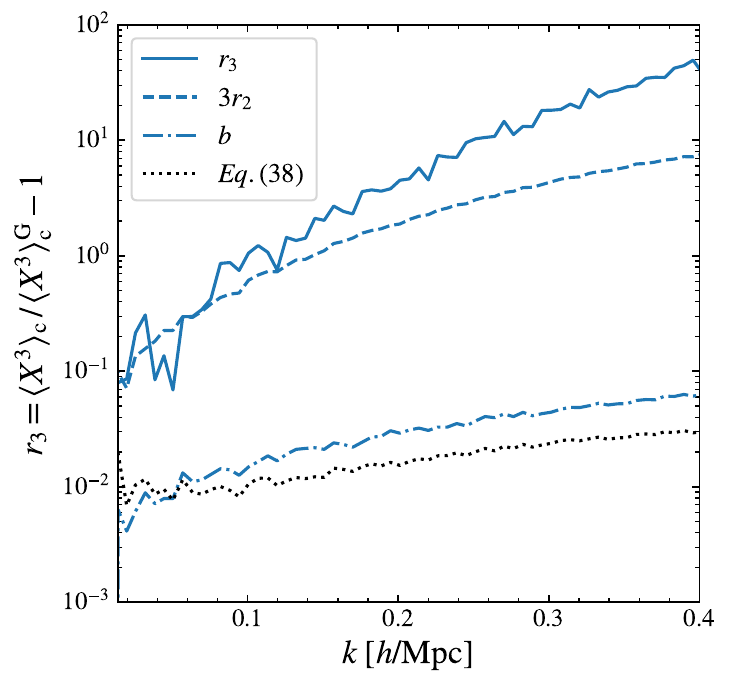}
\caption{Measurement of $r_3$ from the $\Lambda$CDM simulation at z=0 and its different contributions. The solid line shows the total relative difference $r_3$, while the dashed and dot-dashed lines show the trispectrum and bispectrum contribution, $3r_2$ and $b$, respectively. The dotted black line shows a rough estimation of $b$ based on Eq.~\eqref{aproxr3}.}
\label{fig:r3b3}
\end{figure}

By estimating the second- and third-order cumulants from the $\Lambda$CDM simulation at z=0, we can compute $3r_2$ and $r_3$. In addition, we can estimate the equilateral configurations of the bispectrum $B(k)$ on $100\,000$ realisations. It allows us to evaluate the contribution $b$ to the relative excess skewness $r_3$. In Fig.~\ref{fig:r3b3} we compare these three quantities. First, we can see that the overall amplitude of the bispectrum term $b$ is indeed of the order expected from Eq.~\eqref{aproxr3}, which is displayed as a gray dotted line. Second, due to the suppression mechanism described above, the $b$ contribution is more than one order of magnitude smaller than the trispectrum contribution $3 r_2$ at the lowest accessible $k$ and more than two orders of magnitude smaller at $k = 0.2\ \invMpc$. This justifies neglecting the bispectrum term $b$ in the following sections. Thus, we are left with two terms that contribute to the relative excess of skewness $r_3$: the trispectrum term $3r_2$ and the pentaspectrum term which, neglecting $b$, is simply $r_3 - 3r_2$.


\begin{figure}
\includegraphics[width=\linewidth]{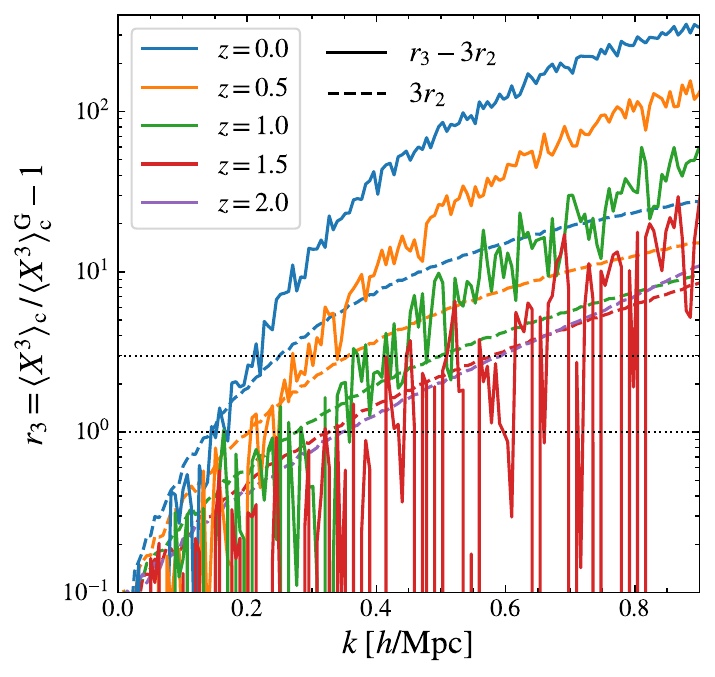}
\caption{Relative excess skewness, $r_3$, made of two contributions: $3r_2$ and $r_3-3r_2$ (respectively in dashed and solid lines). The colours correspond to the different redshifts. For clarity of the figure, we do not show $r_3 - 3r_2$ for $z=2$, as it is compatible with 0. There are two horizontal dotted black lines showing the levels $1$ and $3$.} 
\label{fig:excess_s3}
\end{figure}

In Fig.~\ref{fig:excess_s3} we show how the non-Gaussian contribution to the covariance evolves with redshift in the \covmos~realisations. In this figure $3r_2$ is represented in dashed line so when it crosses the level $3$ (dotted line) it means that the Gaussian and the non-Gaussian part contribute the same amount to the total variance. This is confirming that when the redshift is low ($z=0$), the term $N_k\, k_{\rm F}^3\, \bar T_0/\bar P_0^2$  starts dominating the variance already at $k = 0.25\ \invMpc$ while at high redshift ($z=2$) it happens on smaller scales ($k = 0.6\ \invMpc$). This is a well known result which was already shown by \citet{takahashi_09}.

In addition, in Fig.~\ref{fig:excess_s3} we show the contribution of the two terms $r_3-3r_2$ (pentaspectrum) and $3r_2$ (trispectrum), to $r_3$ (excess skewness) as a function of $k$ for the five redshifts. It appears that at low redshift the term $r_3-3r_2$ starts dominating already at $k > 0.3\ \invMpc$.  However, at high redshift this term fluctuates around $0$ and $r_3$ is rather dominated by $3r_2$. 

Based on the above observations and by expressing the ratio between the total skewness and its prediction for a Gaussian field in terms of $r_2$ and $r_3$ as

\begin{equation}
\frac{S_3^{(0)}}{S_{3,G}^{(0)}} = \frac{1 + r_3}{\left ( 1 + r_2 \right )^{3/2}}\;,
\label{redskew}
\end{equation}
we can obtain approximate expressions in the two regimes (either the pentaspectrum or the trispectrum domination regime) described above. At high redshift, when $r_3$ is dominated by $3r_2$ (trispectrum domination), a simple calculation shows that when $3r_2 \simeq 9$ the ratio in Eq.~\eqref{redskew} deviates from $1$ by only $25$\%. On the other hand, when we consider low redshift and relatively non-linear scales ($k > 0.3\ \invMpc$) we can use the approximation 
\be
r_3 \simeq \frac{N_k^2 k_{\rm F}^6}{2} \frac{\bar Q_0}{P_0^3}\;.
\ee 
As a result, in that regime one can approximate Eq.~\eqref{redskew} as 
\begin{equation}
\frac{S_3^{(0)}}{S_{3,G}^{(0)}} \simeq \frac{\sqrt{N_kk_{\rm F}^3}}{2} \frac{\bar Q_0}{\bar T_0^{3/2}}\;. 
\label{s3approx}
\end{equation}
This shows that the excess of skewness is directly related to the ratio between the shell averaged pentaspectrum and the shell averaged trispectrum to the power of $3/2$. As a result, Eq.~\eqref{s3approx} shows that if this ratio is not well reproduced in the \covmos\ realisations we would not get a realistic excess of skewness.

Thus, one might wonder about the soundness of the \texttt{COVMOS} method to study the non-Gaussian behaviour of the distribution of the power spectrum estimator. To gain confidence in the \texttt{COVMOS} method in terms of higher-order statistics, we conducted a comparison between the \texttt{COVMOS} $\Lambda$CDM simulation at z=0 and a set of $12\,288$ $N$-body simulations using a comparable cosmology, the \texttt{DEUS-PUR} simulations \citep{blot_15}. As demonstrated in Appendix~\ref{validationofcovmos}, our results using \covmos~exhibit a robust agreement with those obtained from $N$-body simulations, up to $k= 0.8\ \invMpc$. This comparison further validates the robustness and accuracy of the \texttt{COVMOS} method for analysing non-Gaussian features.

\begin{figure}
\centering
\includegraphics[width=\linewidth]{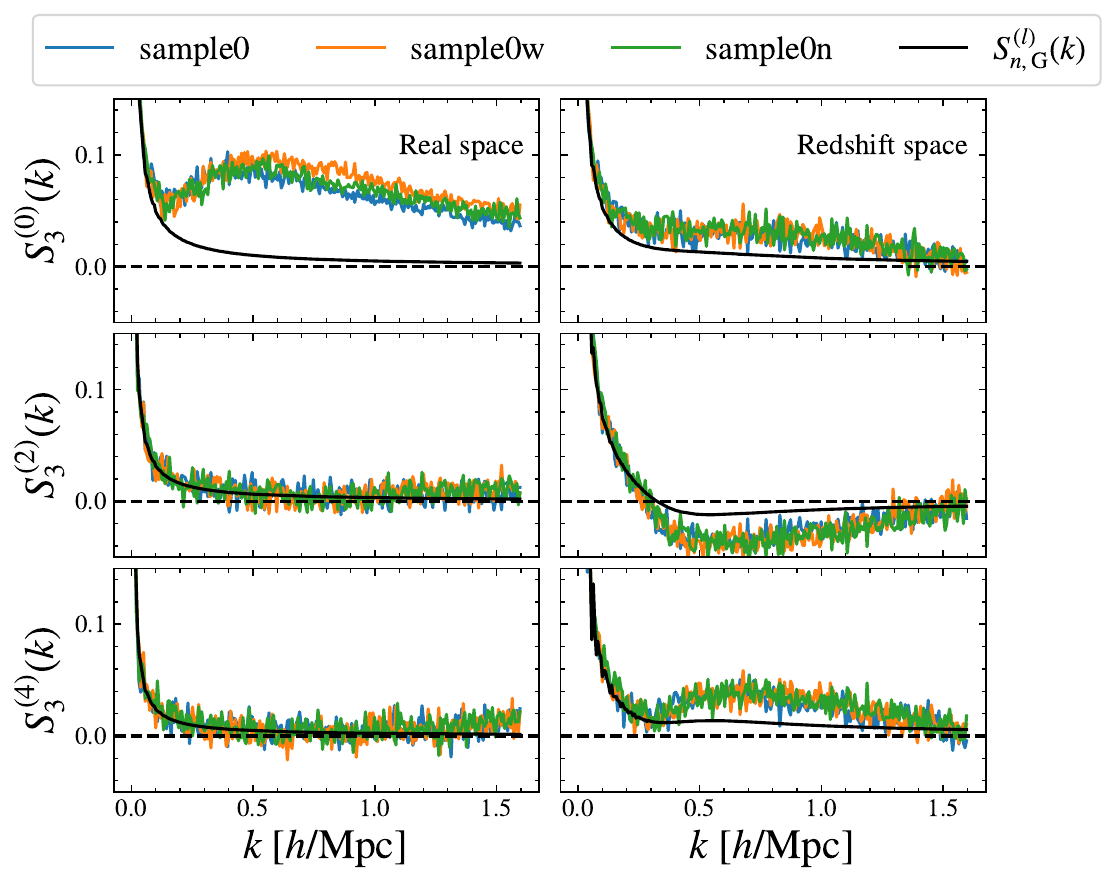}
\caption{ Estimated skewness $S_3^{(\ell)}$ of the distribution of the power spectrum multipoles (from top to bottom panels $\ell = 0,2,4$) $P^{(\ell)}(k)$  in real (left) and redshift space (right) for the $\Lambda$CDM cosmology and its two extensions, 16nu and  w9p3. In each panel the black line is the skewness expected for a Gaussian field.}
\label{fig:skew_model}
\end{figure}

Since the skewness $S_3^{(\ell)}$ of the estimator of the multipoles of the power spectrum depends explicitly on higher-order correlation functions of the density field we also investigate how it changes when considering different cosmological models characterised by roughly the same power spectrum. In Fig.~\ref{fig:skew_model} we compare the skewness estimated in the three samples, $\Lambda$CDM, 16nu and w9p3, corresponding to three different cosmological models (see Table~\ref{tablesamples}) at $z=0$, both in real and redshift space. We can see that adding massive neutrinos (with 16nu) is not changing the skewness, while modifying the dark energy equation of state is only slightly affecting the large $k$ behaviour of the real space monopole. In redshift space all multipoles skewness are unchanged by varying cosmology. It is worth noticing that all three models exhibit similar clustering amplitudes, as evidenced by their power spectra presented in \citet{parimbelli_22}. 

\subsection{Effects of the estimator's setting on the data distribution}

In this subsection we investigate how technical details of the power spectrum estimator can impact its statistical distribution. We focus on the real-space distribution, which maximises the excess of skewness.

\subsubsection{Aliasing and mass assignment scheme}

\begin{figure}
\includegraphics[scale=0.6]{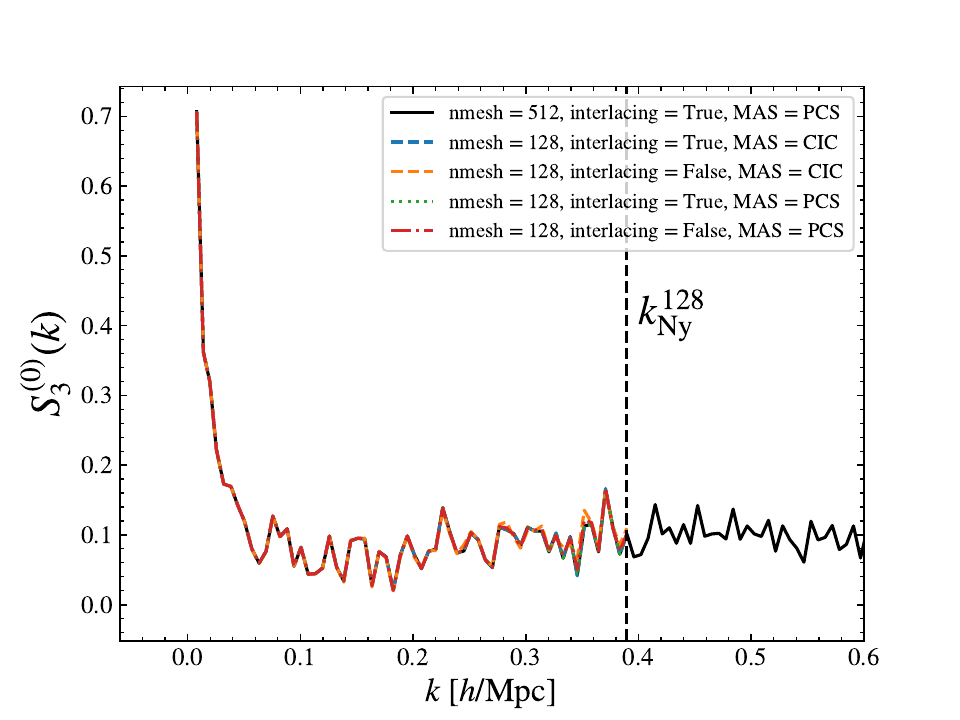}
\caption{Estimated skewness of the power spectrum monopole in real space for different estimator settings on a reduced set of $10~000$ realisations.}
\label{fig:inter_test}
\end{figure}
Despite the soundness of the results presented above, we want to verify their robustness against changes in the choices made for the settings of the estimator. One could think of a possible dependence in the choice of the mass assignment through aliasing effects. Indeed, aliasing is introducing some correlations between modes which might induce some unexpected skewness \citep[][]{baratta_19}. Thus, to ensure that the observed excess of skewness in the power spectrum distribution is indeed cosmological and do not arise from any numerical artefact, we modified and compared different settings of the estimator.

We tested three different settings. The first was (i) the interlacing technique \citep{sefusatti_15}, in which one can choose to reduce aliasing arising when using fast Fourier transforms. The second concerned (ii) the type of mass assignment scheme used to interpolate the particles over the grid before estimating the power spectrum, where we tested two of them: the cloud-in-cell and piecewise cubic spline. The third involved (iii) the precision of the grid, where we chose a standard mesh of $N_\mathrm{mesh} = 512$ points per dimension and a less precise one with $N_\mathrm{mesh} = 128$. 

The results are shown in Fig.~\ref{fig:inter_test}. They clearly demonstrate that, regardless of the choice of these three settings, the skewness of the power spectrum distribution remains the same, including when close to the Nyquist mode $k_\mathrm{Ny}^{N_\mathrm{mesh}} = \pi N_\mathrm{mesh}/L$. This indicates that the type of estimator used does not affect the skewness of the power spectrum.

\subsubsection{Fourier mode binning}
\label{sec_bin}
In galaxy survey analyses, it is common to adapt the power spectrum wave-mode bin width $\Delta k$ to accurately capture the cosmological information across various scales. The goal is to find the optimal $\Delta k$ to balance the need for detailed understanding of the power spectrum behaviour across scales while minimising the noise, the correlations between different bins and the number of bins in the data-vector. For example it is possible to enlarge the binning in $k$ such that one can reduce the size of the data-vector. This is particularly useful to obtain cosmological constraints based on a covariance matrix which is estimated from a finite set of realisations \citep{gouyou_25}. 

We thus want to observe the effect of the bin width on the skewness of the estimator of the power spectrum distribution in light of the analytical formalism presented in Sect.~\ref{sect:theory}. Indeed, we expect that enlarging the bin width will affect the shell averaged poly-spectra entering in the third-order cumulant of the distribution of the power spectrum estimator (see Eq.~\ref{skewnessfull}). In the limit of a Gaussian density field, where the skewness is given by Eq.~\eqref{s30gauss}, the only effect we can expect is that, by increasing the size of the bins, the number of modes within the Fourier shell will also increase, resulting in a decrease of the skewness. However, considering a non-Gaussian density field it is difficult to predict the effect of the binning with simple arguments. As a result, in this section we systematically test the effect of increasing the bin width of the estimated monopole power spectrum in real space.

\begin{figure}
\includegraphics[width=\linewidth]{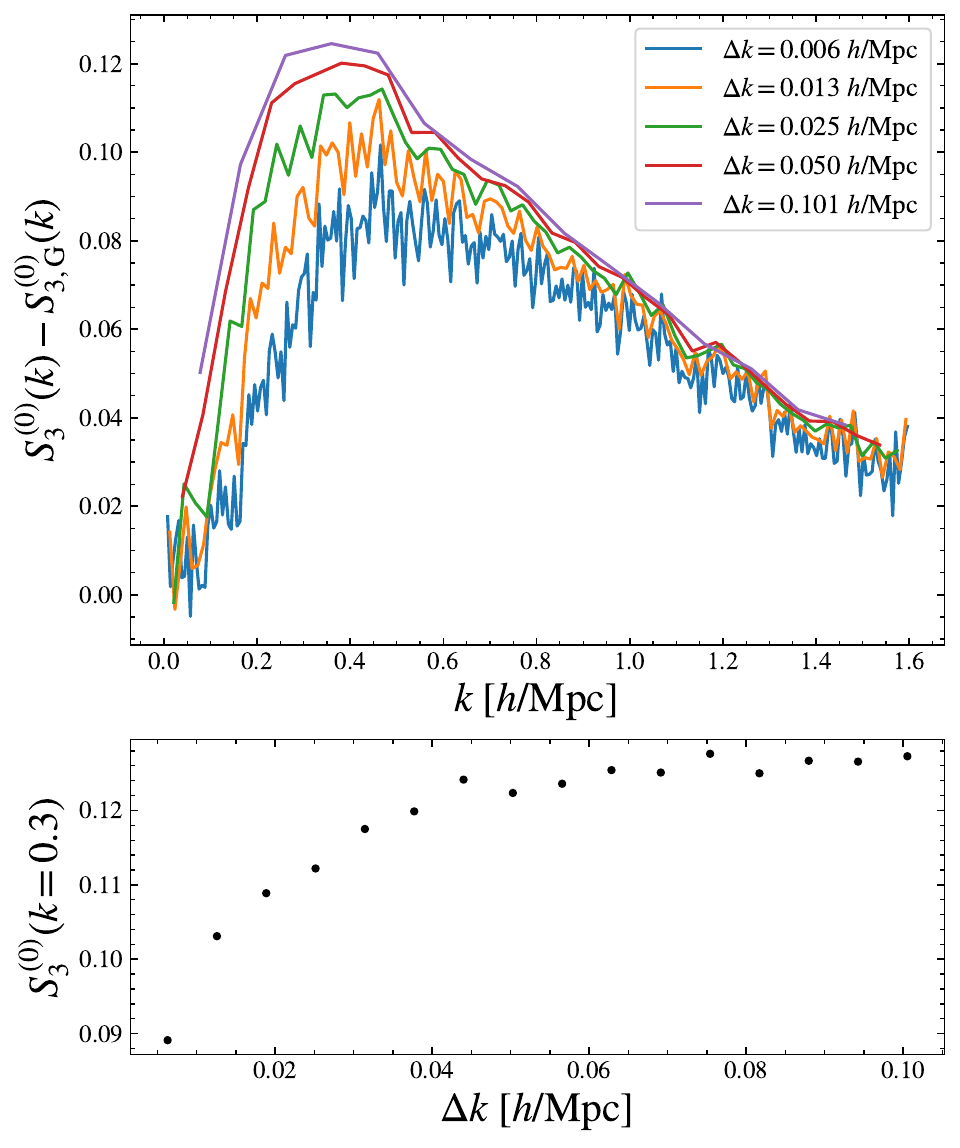}
\caption{Impact of wave mode binning. {\it Top}: Difference between the estimated skewness and the Gaussian field prediction for the real-space power spectrum at $z=0$ for different bin sizes, $\Delta k$. {\it Bottom}: Skewness at $k=0.3\ \invMpc$ with respect to $\Delta k$.}
\label{fig:binning}
\end{figure}

Figure~\ref{fig:binning} displays the excess skewness of the power spectrum distribution with respect to the Gaussian field prediction, at $z=0$ and in real space, for various bin widths, ranging from $\Delta k =k_{\rm F} = 0.006\ \invMpc$ to $\Delta k = 0.1\ \invMpc$. We can see that increasing the bin width produces more excess skewness, especially on intermediate scales, between $k=0.15$ and $0.4\ \invMpc$. In that range we can see that the excess skewness can increase by 66\% going from the reference binning size (i.e. the fundamental frequency) to largest one. On these scales higher-order statistics, and thus the non-Gaussian terms in Eq.~\eqref{skewnessfull}, start to be relevant. On the contrary, for $k>0.55\; \invMpc$ the increase with the binning size is less than $15$\%. Since in this regime the skewness can be approximated with Eq.~\eqref{s3approx}, it means that the ratio between the pentaspectrum and trispectrum to the power $3/2$ is close to be constant when increasing the shell size.

In the lower panel of Fig.~\ref{fig:binning} we can appreciate how the skewness is increasing with the binning size at fixed $k=0.3\; \invMpc$. It appears that it saturates at around 7 times the fundamental frequency. In conclusion, even a drastic increase in the binning size does not increase the skewness of the estimator of the power spectrum more than $50$\%.

\subsection{Effects of survey-related features on the data distribution}

In this subsection, we investigate the impact of characteristic features of a survey on the statistical distribution of the power spectrum estimator. Since we have shown in Sect. \ref{sect:measure} that being in real space maximises the excess of skewness we keep the same setting in the following.

\subsubsection{Sample density}
First, we examine the effect of the number density of objects in the catalogue. Indeed, since shot noise affects all $N$-point correlation functions, it should induce a non-zero trispectrum and pentaspectrum and can thus modulate the expected skewness when the Gaussian contribution is sub-dominant. Our reference catalogue, $\Lambda$CDM at $z=0$, has a number density $\rho = 0.1\ \invMpcc$. We generated a set of three sparser catalogues with densities of $\rho = 0.05$, $0.01$, and $0.001\ \invMpcc$ and for each of them we estimate the skewness of the real-space monopole power spectrum. 

\begin{figure}
\includegraphics[width=\linewidth]{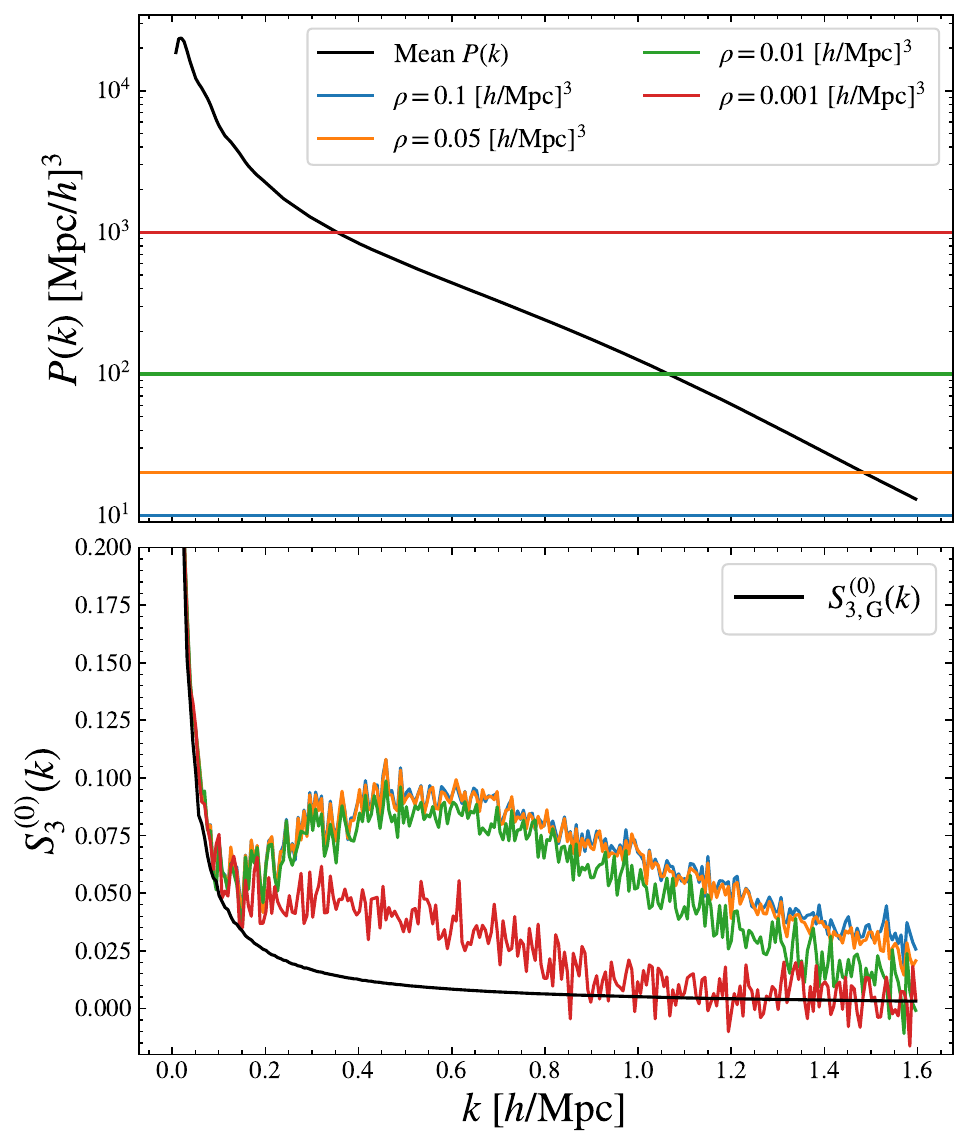}
\caption{Impact of the shot noise. {\it Top}: Mean power spectrum for the reference sample and the different level of shot noise corresponding to each density. {\it Bottom}: Estimated skewness for the different densities considered (coloured lines) and the Gaussian field prediction (black line).  }
\label{fig:skew_sn}
\end{figure}

The top panel of Fig.~\ref{fig:skew_sn} displays the average power spectrum of the reference sample, as well as the shot noise amplitude of the other samples. The middle panel shows the estimated skewness $S_3^{(0)}(k)$ in all samples. From this figure, it is evident that increasing the shot noise level tends to lower the excess skewness. The suppression starts to be effective at the wave mode for which the power spectrum and the shot noise level have the same amplitude. One can notice that when shot noise dominates completely over the signal, no excess skewness becomes detectable. This is in agreement with the expectation. In fact, when the signal is dominated by shot noise we expect the power spectrum to be equal to $P_{\rm SN}=1/\rho$. In addition, the higher-order $N$-point spectra such as the bispectrum, trispectrum, and pentaspectrum are affected in a non-trivial way by the shot noise \citep[see][for the explicit expressions up to the trispectrum]{blot_15}. However, it is fairly intuitive that a pure Poisson distribution will be characterised by $B(\vec k_1, \vec k_2) = P_{\rm SN}^2$, $T(\vec k_1, \vec k_2, \vec k_3) = P_{\rm SN}^3$ and $Q(\vec k_1, \vec k_2, \vec k_3, \vec k_4, \vec k_5) = P_{\rm SN}^5$. Thus, at high $k$ we expect $\langle X^2\rangle_c \simeq k_{\rm F}^3P_{\rm SN}^3$ and $\langle X^3 \rangle_\mathrm{c} \simeq k_{\rm F}^6 P_{\rm SN}^5$, from which it follows that 

\begin{equation}
S_3 \simeq (k_{\rm F}^3P_{\rm SN})^{1/2},
\label{s3_shot}
\end{equation}
when the signal is dominated by shot noise. The Eq.~\eqref{s3_shot} suggests that the skewness should increase with the shot-noise power spectrum $P_{\rm SN}$ which seems contrary to what is observed in Fig.\ref{fig:skew_sn}.  However, considering the most extreme shot-noise case ($\rho=0.001\ \invMpcc$) we expect the skewness to decrease to a value of $0.016$ at large $k$, which is a level too low to be detected in our case.  
The fact that the shot noise reduces the skewness was also observed in \citet{lin_20} in the case of weak lensing two-point correlation functions.

In \citet{EuclidSkyOverview} it is predicted that the total spectroscopic sample of \Euclid will have a density of less than $\rho = 10^{-3}\ \invMpcc$, which is comparable to the highest level of shot noise shown in Fig.~\ref{fig:skew_sn}.

\subsubsection{Survey window function}\label{sect:mask}

\begin{figure}
\includegraphics[width=\linewidth]{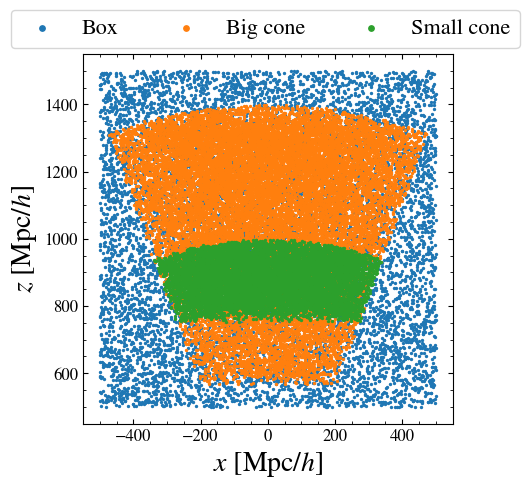}
\caption{Two-dimensional projection of a sample of particles from the three different geometries considered: the periodic box and the big and small cone. }
\label{fig:masks}
\end{figure}

As observed in \citet{upham_21}, another observational aspect that can significantly impact the distribution of two-point statistics is the survey mask. In order to investigate this effect, we carve two arbitrary cone-like geometries within the periodic boxes generated with \covmos. We define two different cones that both have an aperture of semi-angle $\theta=\pi/9$. The big cone has a radial extension of $\left[r_\mathrm{min}, r_\mathrm{max}\right] = \left[600,\ 1\,400\right]\ \mathrm{Mpc}/h$ and the small one $\left[r_\mathrm{min}, r_\mathrm{max}\right] = \left[800,\ 1000\right]\ \mathrm{Mpc}/h$. 
The big and small cones have respectively a volume  of $0.3$ $(\mathrm{Gpc}/h)^3$ and $0.06$ $(\mathrm{Gpc}/h)^3$. The particle number density is the same as the reference value of $\rho = 0.1\ \invMpcc$ that we considered for the periodic box. In Fig.~\ref{fig:masks} we show how the small and big cones are compared with the reference comoving box.

We estimate the power spectrum in each of the $100\,000$ realisations at $z=0$, in real space, for both cone-like geometries with an enclosing box of sised $1000h^{-1}$Mpc. As we are no longer analysing a periodic box we need to take care about the effect of the mask in the estimation of the power spectrum (see \citealt{beutler_17} or GB-2025 for more details). To have an accurate estimation of the survey mask power spectrum we take the average power spectrum from 50 realisations of a uniform random distribution following the cone geometries with 20 times the reference density.

From the distribution of the measured power spectra we can estimate the skewness of its distribution (Eq.~\ref{s3estimator}) in the small and big cones and compare them to the reference periodic box. In addition, we compare these to the Gaussian field prediction. However, if in the case of a periodic box it is fairly straightforward  to get this prediction (see Appendix~\ref{appendixa}), in the case of a masked field calculations become significantly more complicated. We need to be careful about the effect of the integral constraint \citep[IC,][]{demattia_19}, which is expected to propagate into the skewness. 

The IC comes from the fact that the density contrast is defined from the mean number density of objects in a given catalogue. However, we can only estimate it from the same volume that we have at our disposal to estimate two-point statistics. This gives rise to correlation between large- and short-scale modes. Thus, in order to properly compare the skewness estimated in the \covmos~realisations with respect to the Gaussian field case, we generate $N_{\rm G}=100\,000$ realisations of masked Gaussian fields with and without taking into account the IC. We set-up these Gaussian fields, dubbed $\nu(\vec x)$ in the following, such that they have the same power spectrum as the \covmos~realisations.

In the following we explain how we include or not the contribution from IC in these Gaussian realisations. As explained above, the mean number density $\bar n$ that we estimate inside a given mask of volume $V$, as $\bar n = N_{\rm gal}/V$, is different from the true number density $n_0=\rho$. Thus, within the same volume $V$ the estimated density contrast $\delta_{\rm IC}(\vec x)$ is different from the true one $\delta(\vec x)$. The relation between the two can be obtained by noticing that the only quantity which is conserved is the local number density $n(\vec x)$, implying that 

\begin{equation}
\bar n \left [ 1 + \delta_{\rm IC}(\vec x) \right] = n_0 \left [ 1 + \delta(\vec x) \right]\;.
\label{localdens}
\end{equation}
Due to the finite size of the geometry, $\bar n$ fluctuates from one realisation to another such that $\bar n = n_0(1+\epsilon)$, where 

\begin{equation}
\epsilon := \frac{1}{V} \int_V \delta(\vec x)\,\dif^3\vec x\;.
\end{equation}
As a result, one can relate the Gaussian field $\nu(\vec x)$  to the corresponding Gaussian field $\nu_{\rm IC}(\vec x)$ affected by the IC  directly with 

\begin{equation}
\nu_{\rm IC} (\vec x) = \frac{\nu(\vec x) - \epsilon}{1+\epsilon}\;.
\label{nuic}
\end{equation}
We can thus efficiently estimate the skewness of the power spectrum estimated in the Gaussian realisations with and without the presence of the IC. We found that without the IC, one can find an empirical model for the skewness of a Gaussian field in the cone geometries. Indeed, by defining an effective fundamental mode for a given geometry of volume $V$ as $k_{{\rm F},\mathrm{eff}} := 2\pi/V^{1/3}$, one can define an effective number of independent modes in a given $k$-shell as $N_{\rm eff} := 2\pi\, k^2/ k_{{\rm F},\mathrm{eff}}^2$. Empirically we found that $S_{3,{\rm G}}^{(0)} \simeq 2.2/\sqrt{N_{\rm eff}}$ provides a good fit for both volumes we consider. 

\begin{figure}
\includegraphics[width=\linewidth]{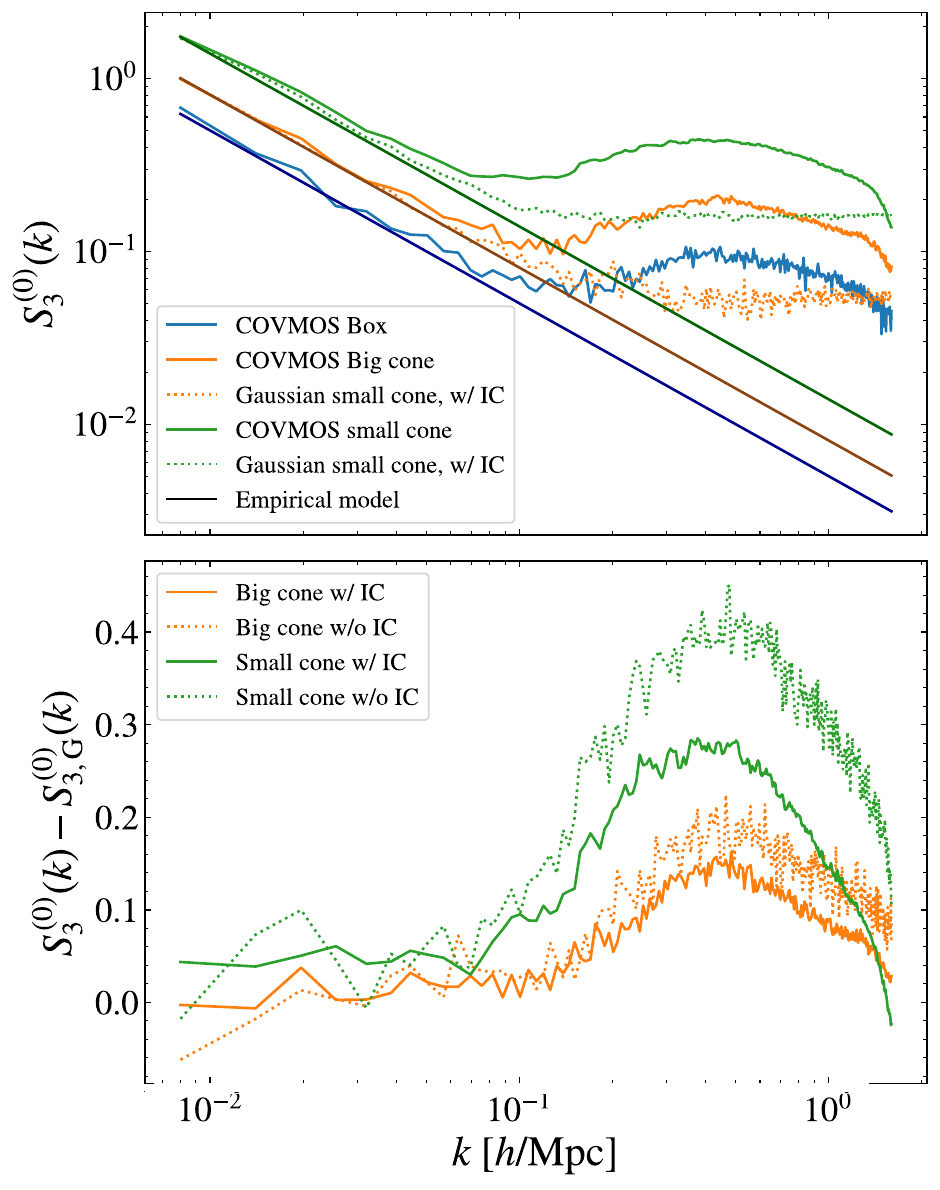}
\caption{Impact of the survey mask. {\it Top}: Estimated skewness in the three geometries from the \covmos~ simulations (light coloured plain lines) and the Gaussian field realisations, where the contribution from IC is not subtracted (dotted lines). We also show our empirical model for the Gaussian field without the IC contribution (dark coloured solid lines). {\it Bottom}: Difference between the skewness estimated in the \covmos~ simulations and the Gaussian realisations with and without IC (light coloured plain and dotted lines, respectively). }
\label{fig:skew_mask}
\end{figure}

In Fig.~\ref{fig:skew_mask} we show the skewness estimated from the \covmos~realisations for the two cones and the periodic box and compare them to the skewness estimated from the Gaussian realisations described above. We can see that the presence of a mask increases the level of skewness at all scales. The smaller is the mask the larger is the increase of the skewness level. This is expected because the application of the mask on the density field correlates Fourier modes. Indeed, this is in a way similar to increasing the size of the $k$-shell (see Sect. \ref{sec_bin}).

Our empirical model, which reproduces the skewness obtained from a Gaussian field without IC is shown in the top panel of Fig.~\ref{fig:skew_mask} with dark solid lines. Comparing this empirical model, which does not include the effect of IC, to the skewness estimated from the Gaussian field with IC (dark solid lines against dotted lines in the top panel of Fig.~\ref{fig:skew_mask}), we observe that on small scales the IC itself is introducing an excess of skewness in the distribution of the power spectrum. We notice that its amount depends on the volume used to estimate the mean number density and that it is  roughly constant at large $k$.   

As a result, to quantify the excess skewness with respect to the Gaussian case, one needs to subtract the Gaussian skewness with IC to the estimated skewness in the \covmos~catalogues. This is shown in the lower panel of Fig.~\ref{fig:skew_mask}, where it appears clearly that when the considered volume is small ($0.06$ [Gpc/$h$]$^3$) it is important to take into account the skewness induced by the IC while when the volume is larger ($0.3$ [Gpc/$h$]$^3$) the absolute contribution from the IC to the total excess skewness becomes subdominant. As a reference, the smallest redshift bin for the \Euclid spectroscopic sample of the first data release (DR1) is expected to be around $1.5\; \Gpcc$, thus the IC should not increase the level of skewness in the \Euclid two-point summary statistics in Fourier space.

\section{Statistical distribution of individual Fourier modes within a shell}
\label{sect:single_mode}

Up to now we studied the distribution of the shell-averaged estimated power spectrum and found that the excess of skewness with respect to a Gaussian distribution originates from the shell-averaged higher-order statistics, such as the trispectrum and pentaspectrum. We now want to understand the origin of this non-Gaussian distribution and whether it is generated by an intrinsic non-Gaussian distribution of each individual Fourier mode. In this section, we are thus interested in the distribution of square modulus of the density contrast  $ X_i := |\delta_{k_i}|^2$ evaluated at each individual wave mode within a shell $k$.

As shown in Appendix~\ref{appendixa}, when $\delta_{\vec k}$ is a Gaussian field then $X_i$ is expected to follow an exponential distribution. We thus, want to understand whether the skewness in the shell averaging process is generated by an intrinsic departure from the expected exponential distribution of each individual $X_i$ or from the correlations between them. To achieve this we focus on the distribution of a sub-sample of modes within four Fourier shells $k=0.2\ \invMpc$, $k=0.3\ \invMpc$, $k=0.4\ \invMpc$, and $k=0.5\ \invMpc$. In each $k$-shell, we randomly select $1000$ square modulus $X_i$, for which we can compute the one-point cumulant moments up to an order of five.  
We can compare these cumulant moments to the predicted ones assuming an exponential distribution (i.e. in the case of a Gaussian field). The latter can be expressed as

\begin{equation}\label{eq:gauss_cum}
    \langle X_i^n\rangle_{\rm c}  = (n-1)!\, \langle X_i\rangle^n\;,
\end{equation}
as shown in Appendix~\ref{appendixa}.

\begin{figure}
\includegraphics[width=\linewidth]{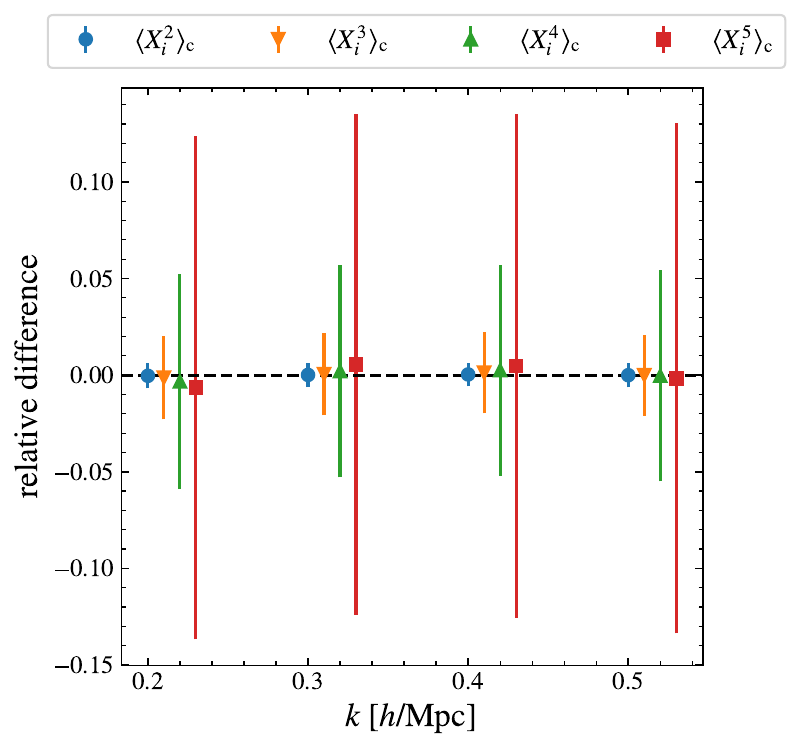}
\caption{Relative difference between the cumulants (up to an order of five) of each individual mode and their predictions in the case of a Gaussian field (see Eq.~\ref{eq:gauss_cum}). The error bars show the dispersion obtained for the $1000$ considered wave modes.}
\label{fig:cumulants}
\end{figure}

In Fig.~\ref{fig:cumulants} we display the relative deviation between the estimated cumulant moments of $X_i$ with respect to their prediction of Eq.~\eqref{eq:gauss_cum}. Since for each shell we have $1000$ square modulus $X_i$ we only show the mean relative deviation (computed over the $1000$ modes) and the corresponding dispersion over the considered shell. We can see a very good agreement between the estimation and its prediction in the case of an exponential distribution.

Despite the fact that the first five cumulant moments are well reproduced when fixing the first moment to the expectation value evaluated over the $100\,000$ realisations, one could argue that this does not prove that the full distribution is well represented by an exponential distribution. Thus, we conducted a Kolmogorov--Smirnof (KS) test to compare each distribution of the $4\times 1000$ wave modes with an exponential distribution. For each of the $4\times 1000$ wave modes, we generate a set $100\,000$ values extracted from an exponential distribution with the estimated mean value and we perform the KS test to compare the true distribution and the exponential one. It turns out that the lowest probability assessing the inconsistency between the observed and expected distribution over all the tested modes is $0.73$. It means that for none of the $4000$ modes at our disposal we can reject the hypothesis that the distribution is exponential without having a very high probability of drawing a wrong conclusion.
Thus, it appears that the square modulus at each measured wave mode is following with a great accuracy an exponential distribution.

In conclusion, this means that despite the fact that the density field is non-Gaussian, the distribution of the square modulus of the Fourier space density field is following the distribution which would be expected if the field were Gaussian. This is a clear indication that the main source of the excess skewness of the power spectrum distribution is due to the correlation between wave modes.

\section{Conclusions}\label{sect:conclu}

Given the level of precision of the cosmological constraints the \Euclid mission aims at providing, the reliability of common assumptions that are usually made on the likelihood function must be checked to the same level of precision. One crucial assumption that was studied in detail in this paper is to consider that two-point statistics follow a Gaussian distribution. 

We could test this assumption and compare it to our theoretical expectations with great precision thanks to the \covmos~method, which has been proven to be suitable to study the covariance matrices of two-point statistics \citep{baratta_19, baratta_22, gouyou_25} and the $N$-point distribution of two-point statistics in the present work (see the comparison with the \texttt{DEUS-PUR} simulations in Appendix~\ref{validationofcovmos}).

We first derived the analytical expressions of the second- and third-order cumulants of the power spectrum distribution in terms of the $N$-point functions of the density field. This allowed us to directly put in evidence of the relation between the non-Gaussianity of the density field and the non-Gaussianity of the power spectrum distribution and isolate the competing terms contributing to this distribution (see Eqs.~\ref{usualvar} and \ref{skewnessfull}). In order to understand which terms were dominating depending on the scales, we derived the analytical expressions of the skewness and kurtosis of the multipoles of the power spectrum in both real and redshift space assuming a Gaussian field (see Eqs.~\ref{s30gauss} to \ref{s44gauss} and Appendix~\ref{appendixa}).

To compare these approximate predictions to their equivalent estimations for a more realistic non-Gaussian field, we generated several simulation samples, each containing $100\,000$ \covmos~realisations of the matter density field, while varying the redshift and the cosmological model both in real and redshift space. This allowed us to estimate the skewness and kurtosis of the distribution of the power spectrum monopole in real space and confirm, as was already observed in \citet{takahashi_09} and \citet{blot_15}, the presence of an excess of skewness on non-linear scales ($k>0.1\ \invMpc$ at $z=0$). In addition, for the first time, we also estimated the skewness and kurtosis of the power spectrum multipoles in redshift space. We observed that the presence of RSD transfers the excess skewness of the monopole in real space to the multipoles but significantly reduces its overall amplitude (by roughly a factor of ten).

Based on our predictions and the estimation of the skewness in the \covmos~samples, we could demonstrate that the bispectrum contribution to the skewness (see Eq.~\ref{bdef}) is negligible because only a small fraction of the triangular configurations contribute to $\langle X^3 \rangle_\mathrm{c}$. This led us to understand that on intermediate scales, the excess skewness is dominated by the trispectrum contribution, while on small scales it is dominated by the pentaspectrum of the matter density field.

We also extended this study to other cosmological models, including massive neutrinos and a time-varying dark energy equation of state. We have shown that varying the cosmological model does not have a significant impact on the skewness.

We then varied different settings in the estimation of the power spectrum to gauge their effects on its skewness. We observed that the aliasing due to the use of fast Fourier transforms has no impact on the skewness. However, varying the size of the $k$-shells of the power spectrum has a significant impact on the skewness, as increasing their size produces an increase in the excess skewness of up to 20\%. 

We also explored how the skewness was impacted by the number density of objects, which is a feature that is specific to a given survey. We saw that increasing the level of shot noise has the effect of reducing the skewness of the power spectrum on scales where the power spectrum is dominated by the shot noise. For a level of shot noise similar to the one expected for the \Euclid spectroscopic sample, the skewness is greatly reduced (by around 50\% for $k \sim 0.4 \; \invMpc$, which are the most impacted scales).

However, all of this was done while assuming a periodic universe in a box, which is far from the reality of the data that will be obtained with \Euclid (or any other survey). Indeed, we found that restricting the density field to a cone-shaped survey mask significantly increases the skewness of the power spectrum in two different ways. Firstly, reducing the surveyed volume necessarily reduces the number of available Fourier modes, but it also correlates them, thus increasing the level of skewness at all scales. Secondly, the presence of the IC has a specific impact on the excess skewness on small scales, increasing it for smaller volumes.

Finally, in an attempt to understand the origin of the skewness of the power spectrum, we looked at the statistical distribution of each individual wave mode, $X = k_{\rm F}^3\,|\delta_{\vec k}|^2$. We found, with a high level of accuracy, that they conserve an exponential distribution ($\theta^{-1}\,{\rm e}^{-X/\theta}$,  with parameter $\theta = k_{\rm F}^3\,\langle |\delta_{\vec k}|^2\rangle$) no matter their configurations. This led us to conclude that the extra skewness observed on small scales appears because of the correlation between wave modes within a $k$-shell.

To conclude, it is clear from this work that the distribution of two-point statistics is not Gaussian, and we understand the reasons for that. In particular, the shot noise, the scale binning, and the survey mask are factors that can significantly enhance its skewness. However, it is still unclear whether the assumption of a Gaussian likelihood, given non-Gaussian distributed data, would bias parameter inference for a dataset similar to what will be obtained with \Euclid. This is explored in our companion paper GB-2025, where we show that these non-Gaussian features do not introduce a significant bias in cosmological parameter estimation for Euclid-like datasets.

\begin{acknowledgements}
  
\AckEC
The project leading to this publication has received funding from Excellence Initiative of Aix-Marseille University -A*MIDEX, a French "Investissements d’Avenir" programme (AMX-19-IET-008 -IPhU). 
The DEMNUni-cov simulations were carried out in the framework of “The Dark Energy and Massive Neutrino Universe covariances" project, using the Tier-0 Intel OmniPath Cluster Marconi-A1 of the Centro Interuniversitario del Nord-Est per il Calcolo Elettronico (CINECA). We acknowledge a generous CPU and storage allocation by the Italian Super-Computing Resource Allocation (ISCRA) as well as from the coordination of the “Accordo Quadro MoU per lo svolgimento di attività congiunta di ricerca Nuove frontiere in Astrofisica: HPC e Data Exploration di nuova generazione”, together with storage from INFN-CNAF and INAF-IA2.
\end{acknowledgements}

\bibliography{Euclid}

\begin{appendix}

\section{Skewness and kurtosis for the multipoles}
\label{appendixa}

In this appendix, we derive the expressions of the estimation of the skewness $S_3^{(\ell)}$ and kurtosis $S_4^{(\ell)}$ of the estimator of the power spectrum multipoles. We define the estimator $X$ as a shell average at a fixed $k$ modulus as

\begin{equation}
X := \frac{2\ell+1}{N_k} \sum_{i=1}^{N_k} |\delta_{\vec k_i}|^2 \lcal_\ell(\mu_i)\;,
\label{pkestimator}
\end{equation}
where the sum runs over all modes contained in a given shell and $\mu_i$ is the cosine of the angle between the wave vector $\vec k$ and the line-of-sight (along the $z$ direction of the periodic box). Defining the variable $X_i:= |\delta_{\vec k_i}|^2$ and the coefficients $a^{(\ell)}_i:= k_{\rm F}^3(2\ell+1)\lcal_\ell(\mu_i)/N_k$, one can express the estimator as a weighted average over the random variable $X_i$,

\begin{equation}
X = \sum_{i=1}^{N_k} X_ia_i^{(\ell)}\;.
\end{equation}
Since the variable $X_i$ is made of the sum of the square of the real and imaginary part of the Fourier modes $\delta_{\vec k_i}$, then it means that it follows an exponential distribution $\pcal$ \citep{fisher_93} of parameter $\theta_i:= P(\vec k_i)$, and thus

\begin{equation}
\pcal(X_i) = \frac{1}{\theta_i} {\rm e}^{-X_i /\theta_i}\;. 
\end{equation}
Its associated moment generating function $\mcal(t_i):= \langle {\rm e}^{t_iX_i}\rangle$ is therefore given by 

\begin{equation}
\mcal(t_i) = \frac{1}{1-\theta_it_i}\;, 
\end{equation}
and the moment generating function of $Y_i:= X_ia_i^{(\ell)}$ is given by $\mcal_Y(t_i) = \mcal\left (a_i^{(\ell)}t_i\right)$. 
Since we assume that each $X_i$ variable is independent from each other, the same property holds for the $Y_i$ variable, thus one can express the joint $N_k$-point moment generating function of the $N_k$ variables as

\begin{equation}
\mcal_Y(t_1, ..., t_{N_k}) = \prod_{i=1}^{N_k} \frac{1}{1-\theta_ia_i^{(\ell)} t_i}\;.
\end{equation}
Thanks to the properties of the moment generating function, one can express the moment generating function of the sum of $Y_i$ which is the variable $X$ as $\mcal_X(t) = \mcal_Y(t, ..., t)$ and its corresponding cumulant generating function $\ccal(t):= \ln\{ \mcal_X(t) \}$ as

\begin{equation}
\ccal (t) = - \sum_{i=1}^{N_k} \ln \left ( 1 - \theta_ia_i^{(\ell)} t \right )\;. 
\end{equation}
One can take the $n$-th derivative of the above expression to obtain 

\begin{equation}
\ccal^{(n)}(t) = (n-1)! \sum_{i=1}^{N_k} \frac{\left (\theta_i a_i^{(\ell)}\right )^n}{\left (1-\theta_ia_i^{(\ell)}t\right )^n}\;, 
\end{equation}
from which it is straight forward to express the $n$-th order cumulant moments $\langle X^n \rangle_{\rm c}  = \ccal^{(n)}(t=0)$ as 

\begin{equation}
\langle X^n\rangle_{\rm c}  = (n-1)! \sum_{i=1}^{N_k} \left (\theta_ia_i^{(\ell)}\right )^n\;.
\label{momcum}
\end{equation}
Since the skewness $S_3$ and kurtosis $S_4$ are defined as 

\begin{equation}
S_n := \frac{\langle X^n\rangle_{\rm c} }{{\langle X^2\rangle_{\rm c} }^{n/2}}\;, 
\end{equation}
we need to evaluate expression (\ref{momcum}) for $n={2, 3, 4}$. Under some further hypothesis, one can obtain a closed analytical form. Assuming that the discrete sum over the shell can be approximated with its continuous limit, it follows that

\begin{equation}
\langle X^n\rangle_{\rm c}  \simeq \frac{(n-1)!}{N_k^{n-1}}\frac{(2\ell+1)^n}{2}\int_{-1}^{1} P^n(\vec k) \lcal_\ell^n(\mu)\dif \mu\;.
\label{cumulgauss}
\end{equation}
The above equation can be approximated when assuming that we are in real space, thus the power spectrum depends only on the modulus $k$ of the wave vector $\vec k$: 

\begin{equation}
\langle X^2\rangle_{\rm c}  \simeq (2\ell+1) \frac{P^2(k)}{N_k}\;, 
\end{equation}
which is the usual expression for the variance of the power spectrum multipole estimator. For the third-order cumulant moment, we get 

\begin{equation}
\langle X^3\rangle_{\rm c}  \simeq (2\ell+1)^3 2\frac{P^3(k)}{N_k^2}
\left ( \begin{array}{ccc}
\ell \! & \! \ell \!& \! \ell \\
0 \!& \!0 \! & \!0
\end{array}
\right )^2\;, 
\end{equation}
where we see the Wigner $3$-j symbols, and finally for the fourth order, 

\begin{equation}
\langle X^4\rangle_{\rm c}  \simeq (2\ell+1)^4 6\frac{P^4(k)}{N_k^3} \sum_{n=0}^{2\ell} (2n+1)
\left ( \begin{array}{ccc}
\ell\! & \! \ell \!& \!n \\
0 \!& \! 0 \!& \!0
\end{array}
\right )^4\;. 
\end{equation}
This allowed us to express the skewness and kurtosis in the case of a Gaussian field as 

\begin{equation}
S_{3, \rm{G}}^{(\ell)} \simeq \frac{2}{\sqrt{N_k}} (2\ell+1)^{3/2}
\left ( \begin{array}{ccc}
\ell\! & \!\ell\! & \!\ell \\
0 \!&\! 0\! &\! 0
\end{array} \right )^2\;
\end{equation}
and

\begin{equation}
S_{4, \rm{G}}^{(\ell)} \simeq (2\ell+1)^2 \frac{6}{N_k} \sum_{n=0}^{2\ell} (2n+1)
\left ( \begin{array}{ccc}
\ell \!& \!\ell \!& \!n \\
0 \!&\! 0 \!&\! 0
\end{array}
\right )^4\;. 
\end{equation}

\section{Validation of the \texttt{COVMOS} method against the \texttt{DEUS-PUR} $N$-body simulations}
\label{validationofcovmos}

\begin{figure*}
\includegraphics[width=\linewidth]{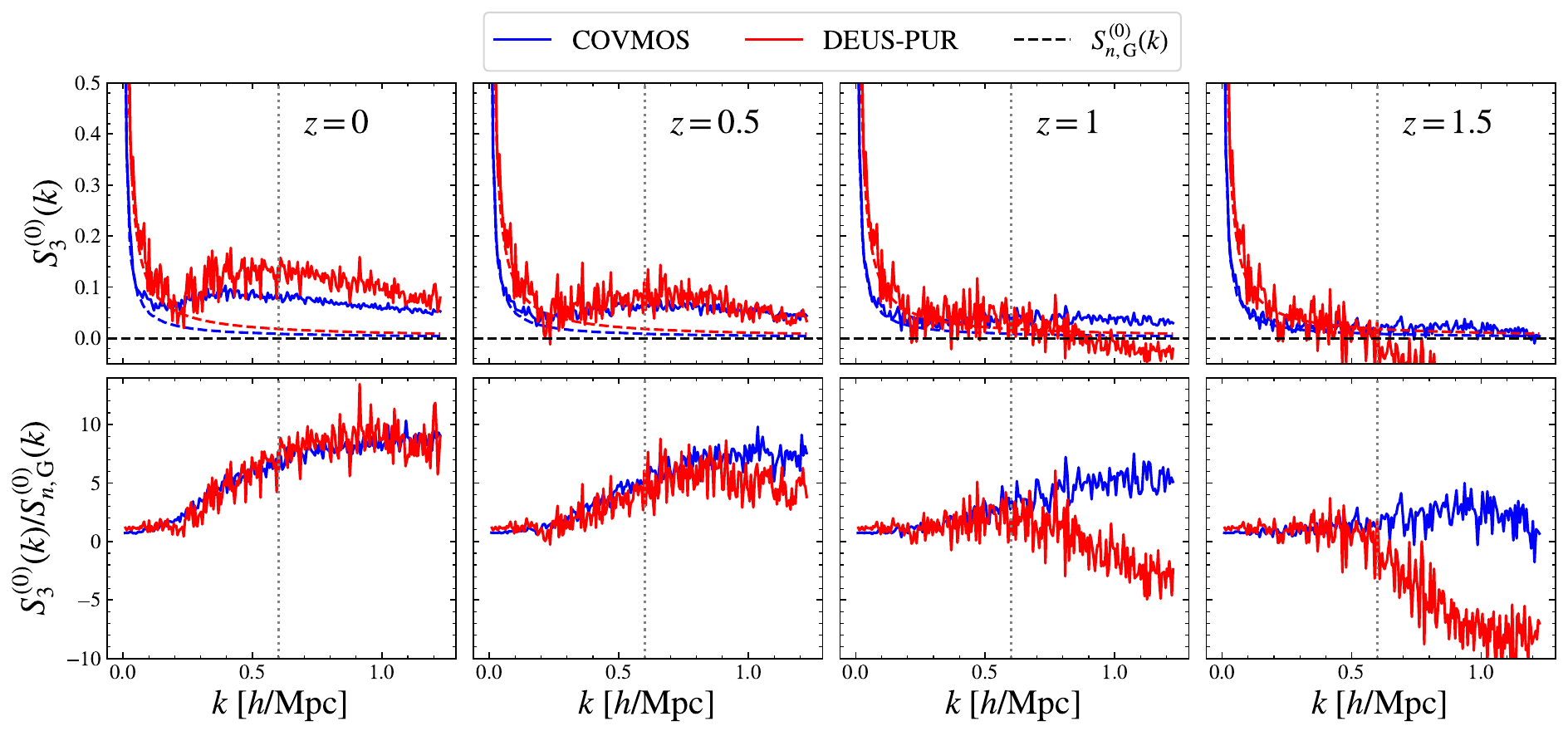}
\caption{Comparison between \covmos\ and \deus . \textit{Top}: Skewness of the distribution of the monopole of the real-space matter power spectrum (\texttt{COVMOS} and \texttt{DEUS-PUR}) in solid blue and red lines. The dashed line represents the corresponding predictions in the case of a Gaussian density field, computed using Eq.~\eqref{s30gauss}.  \textit{Bottom}: Ratio of the estimated skewness to the Gaussian field prediction, following the same colour code.}
\label{fig:comparison_COVMOS_DEUS}
\end{figure*}

Here, we compare the results from the \covmos~realisations to those from the \deus~$N$-body simulation suite. The latter consists of 12\,288 simulations with a volume of $656$~(Mpc/$h$)$^3$ containing $256^3$ tracer particles. All simulations have a $\Lambda$CDM cosmology characterised by the following set of parameter values $(h,\Omega_{\rm m} h^2, \Omega_{\rm b} h^2, n_{\rm s}, \sigma_8) = (0.72,0.1334,0.02258,0.963,0.801)$. Using these simulations, \citet{blot_15} have estimated the skewness of the monopole power spectrum estimator. We use their estimates to validate the skewness of the power spectrum we have obtained from the $\Lambda$CDM \covmos~realisations at $z=0$, 0.5, 1 and 1.5. It is important to remark that the \deus~estimates have been obtained with a binning in $k$ that is set to half the fundamental frequency of the \deus~simulation box. Thus, for a given value of $k$, the number of independent modes may differ from the \covmos~case. 

In Fig.~\ref{fig:comparison_COVMOS_DEUS} we show the estimated skewness $S^{(0)}_{3}(k)$ at $z=0,0.5,1$, and $1.5$ from the \covmos~samples and from \deus. We also show the Gaussian predictions for each case and the corresponding ratios, which exhibit the excess of non-Gaussianity due to the non-linearity of the matter density field. In plotting the excess non-Gaussianity of the \covmos~sample we have rescaled the ratio by a factor of $\sqrt{2}$, as we explain later in the text. 

First of all, we can see that for each case the estimated skewness follows the Gaussian density field prediction on large scales. At $z=0$ this occurs up to $k = 0.2\ \invMpc$ and extends up to $k = 0.5\ \invMpc$ at $z=1.5$. At smaller scales, we can see an excess of non-Gaussianity due to the on-set of non-linearities that propagates from small to large scales for decreasing redshifts. It is worth noticing that in the case of the \deus~results, the estimated skewness drops at large $k$, with a greater effect at large redshifts. A similar trend can also be seen in the case of the \covmos~sample, though with a significant smaller amplitude. This is a numerical simulation resolution effect. In the case of the \deus~simulations, the effect is mitigated at smaller redshifts, where the resolution of the force calculation at small scales is improved by the adaptive-mesh refinement scheme. Indeed, if we conservatively consider only modes corresponding to scales larger than twice the coarse cell of the \deus~simulations, $k\lesssim k_{\rm Ny}/2 = 0.6\ \invMpc$, we are granted that the estimate skewness is free of such numerical systematic errors. The \covmos~samples, on the other hand, have been generated using power spectra estimated from the \demcov~simulations that have greater spatial and mass resolution than the \deus~runs with a Nyquist frequency $k_{\rm Ny}=3\ \invMpc$. That is why such an effect is much smaller than in the case of the \deus~estimates.

Yet, before comparing the excess non-Gaussianity of the \covmos~samples to that from \deus, we should consider the effect of the different binning in $k$ and different volume of the simulations. For this, we can use the approximation of the ratio of the excess skewness given by  Eq.~\eqref{s3approx} since we are looking at non-linear scales ($k\gtrsim 0.2\ \invMpc$ at $z=0$). Indeed, given a binning $\Delta{k}$, we have that $N_{k}\simeq 2\pi\, k^2\, \Delta{k}/k_{\rm F}^3$, thus  Eq.~\eqref{s3approx} reads as

\begin{equation}
\frac{S^{(0)}_{3}}{S^{(0)}_{3,{\rm G}}}\simeq \frac{k\sqrt{2\pi\Delta{k}}}{2}\frac{\bar{Q}_0}{\bar{T}^{3/2}}\;.
\end{equation}
This shows that the simulation volume does not affect that excess skewness, only the binning in $k$. Given that the binning adopted for the \covmos~samples is twice as large as that of \deus, the corresponding excess skewness differs by a factor of $\sqrt{2}$. In the bottom panels of Fig.~\ref{fig:comparison_COVMOS_DEUS}, we have shown the excess skewness of the \covmos~samples rescaled by a factor of $\sqrt{2}$ which agrees well with the estimates from \deus~over the range of non-linear scales which are unaffected by resolution effects. This validates the \covmos~analysis and demonstrates that such synthetic samples are able to reproduce the correct level of excess skewness expected from full $N$-body simulations.

\end{appendix}

\end{document}